\RequirePackage{fix-cm}
\documentclass[twocolumn,epjc3]{svjour3}     

\smartqed  
\usepackage{graphicx}
\usepackage{amsmath}
\usepackage{cuted}
\usepackage{bm}
\usepackage{color}
\DeclareMathOperator{\arcsinh}{arsinh}
\DeclareMathOperator{\atan}{arctan}

\begin{document}

\title{Semiclassical two-step model for ionization by a strong laser pulse: Further developments and applications.}\thanks{Grants or other notes
}


\author{N. I. Shvetsov-Shilovski}

\institute{Institut f\"{u}r Theoretische Physik, Leibniz Universit\"{a}t Hannover, D-30167, Hannover, Germany}
              

\date{\today}

\maketitle

\begin{abstract}
We review the semiclassical two-step model for strong-field ionization. The semiclassical two-step model describes quantum interference and accounts for the ionic potential beyond the semiclassical perturbation theory. We discuss formulation and implementation of this model, its further developments, as well as some of the applications. The reviewed applications of the model include strong-field holography with photoelectrons, multielectron polarization effects in ionization by an intense laser pulse, and strong-field ionization of the hydrogen molecule.  
\end{abstract}

\section{Introduction}
Strong-field physics studies phenomena arising from the interaction of strong laser pulses with atoms and molecules. The most well-known examples of these highly nonlinear phenomena are above-threshold ionization \\(ATI), formation of the high-energy plateau in the electron energy spectrum (High-order ATI), generation of high-order harmonics (HHG) and nonsequential double ionization (NSDI), see Refs.~\cite{DeloneBook2000,BeckerRev2002,MilosevicRev2003,FaisalRev2005,FariaRev2011} for reviews. Both experimental and theoretical approaches used to analyze these processes are constantly being improved. The vast majority of the modern theoretical methods used in strong-field physics are based on the strong-field approximation SFA \cite{Keldysh1964,Faisal1973,Reiss1980}, the direct numerical solution of the time-dependent Schr\"{o}dinger equation (see Refs.~\cite{Muller1999,Bauer2006,Madsen2007,Patchkovskii2017} and references therein), and the semiclassical models applying classical mechanics to describe the electron motion in the continuum. The widely known examples of the semiclassical models are the two-step model \cite{Linden1988,Gallagher1988,Corkum1989} and the three-step model \cite{Krause1992,Corkum1993}.

In the SFA ionization is described as a transition from an initial state unaffected by the laser field to a Volkov state, i.e., the wave function of an electron in an electromagnetic field. Therefore, the SFA neglects the intermediate bound states and the Coulomb interaction in the final state. The SFA provides the illustrative physical picture of many strong-field phenomena and often allows for the analytic solutions. Nevertheless, the approximations used in the SFA are strong enough and may sometimes lead to wrong results. The widely known example is the fourfold symmetry of the photoelectron angular distributions in the elliptically polarized field predicted by the SFA \cite{PPT2}. In contrast to this, the experimental angular distributions show only the inversion symmetry: They are asymmetric in any half of the polarization plane \cite{Bashkansky1988}. The theoretical studies \cite{Basile1988,Lambropoulos1988,Muller1988,Krstic1991,Jaron1999,Manakov2000} have shown that the fourfold symmetry of the angular distributions is a direct consequence of neglecting the effect of the Coulomb potential on the electron motion in the continuum.  

In most cases the direct numerical solution of the TDSE provides a good agreement with the experimental results. However, it is often difficult to understand the physical mechanism of the phenomena under study with only the numerical wave function. What is also important, the capabilities of modern computers are not unlimited. One of the most prominent examples is the strong-field ionization of molecules. The solution of the TDSE in three spatial dimensions is possible only for the simplest molecules and with selection of the most relevant degrees of freedom \cite{Palacios2006,Saenz2014}. Indeed, ionization of a molecule by an intense laser pulse is much more complicated than ionization of an atom. This is because of the existence of additional degrees of freedom (nuclear motion), the associated time scales, and the complex shape of the electronic orbitals. 
For typical laser paremeters used in experiments nuclear motion should be treated on an equal footing with the processes induced by a strong laser field. 
Simultaneously, the rich nuclear structure of molecules results in orbitals of diverse symmetries. 

Although the first semiclassical model (i.e., the two-step model) was formulated in 1988-1989 \cite{Linden1988,Gallagher1988,Corkum1989}, the \\trajectory-based models are still widely used for description of various strong-field phenomena. This is due to a number of important advantages characteristic to the semiclassical approaches. The semiclassical models provide a great insight into strong-field processes. They allow to reveal the specific mechanism responsible for the process under investigation, as well as visualize it using classical trajectories. This point needs to be discussed in more details. 

In the ATI an electron absorbs more photons than necessary for ionization. The studies of the ATI have revealed that the majority of the ionized electrons do not experience hard recollisions with their parent ions. These electrons are referred to as direct electrons. They contribute to the low energy part of the ATI energy spectrum $E<2U_{p}$, where $U_{p}=F_{0}^2/4\omega^2$ is the pomderomotive energy. Here, in turn, $F_{0}$ and $\omega$ are the amplitude and the frequency of the laser field (atomic units are used throughout the paper). The two-step model allows to describe the spectrum of the direct electrons. In the first step of the this model an electron tunnels out of an atom. In the second step it moves along a classical trajectory in the laser field towards a detector.

There are also rescattered electrons that are driven back by the laser field to their parent ions. Upon their returns the rescattered electrons scatter from the parent ions by large angles close to $180^{\circ}$. These electrons form the high-energy plateau of the ATI spectrum. The rescattering scenario provides the basis for an understanding of the HHG and NSDI. Indeed, the returning electron can recombine to the parent ion and as the result of the recombination a high-frequency photon (harmonic) radiation is emitted. Alternatively, if the energy of the scattered electron is sufficient enough, it can release the second electron from the ion, e.g., by impact ionization. The three-step model comprises the interaction of the rescattered electron with the parent ion as the third step. As the result, the three-step model provides the qualitative description of the rescattering-induced processes.

The three-step model explained a number of features revealed in the studies of the high-order ATI, HHG, and NSDI: the cutoffs in high-order ATI spectrum \cite{Paulus1994} and HHG \cite{Krause1992,Lewenstein1994}, the maximum angles of the angular distributions of ionized electrons \cite{Paulus1994a}, the characteristic recoil ion momenta in NSDI \cite{Faria2003,Milosevic2004}, etc. Originally the two-step and the three-step models did not account for the effect of the ionic potential on the electron motion in the continuum. The inclusion of the ionic force in the Newton's equation of motions allowed to uncover the Coulomb focusing effect \cite{BrabecIvanov1996}, study the Coulomb cusp in the angular distributions of the photoelectrons \cite{Dimitriou2004}, investigate the low-energy structures in ionization by the strong midinfrared pulses \cite{Quan2009,Liu2010,Yan2010,Kastner2012,Lemell2012,Lemell2013,Wolter2014,Becker2014,Dimitrovski2015} (the so-called "ionization surprise" observed for the first time in experiment \cite{Blaga2009}), explore the nonadiabatic effects in ionization by intense laser pulses (see, e.g., Refs.~\cite{Boge2013,Hofmann2014,Geng2014}), etc. 

The trajectory-based simulations are often (although not always) computationally less expensive than the solution of the TDSE. Furthermore, for some strong-field processes the semiclassical simulations are presently the only feasible approach. The most well-known example of such process is the NSDI of atoms by circularly \cite{Uzer2010} or elliptically polarized pulses \cite{Shvetsov2008,XuWang2009,Hao2009}, as well as the NSDI in molecules \cite{Agapi2011}. Therefore, further development of the semiclassical approaches to strong-field phenomena is an important objective. 

Until recently the trajectory-based models were not able to describe quantum interference effects. However, a significant progress along these lines has been made in the last decade. The trajectory-based Coulomb SFA (TCSFA) \cite{Yan2010,Yan2012}, the quantum trajectory Monte Carlo model (QTMC) \cite{QTMC}, the semiclassical two-step model (SCTS) \cite{Shvetsov2016}, and the Coulomb quantum orbit strong-field approximation (CQSFA) \cite{Faria2017a,Faria2017b,Faria2018aa,Faria2018a,Faria2018b} (see Ref. \cite{Faria2015} for the foundations of the CQSFA approach) are recent trajectory-based models that are capable to reproduce interference structures in photoelectron momentum distributions of the ATI process. These models assign certain phases to classical trajectories, and the contributions of different trajectories leading to the same final momentum are added coherently.

The TCSFA is an extension of the CCSFA \cite{Popruzhenko2008a,Popruzhenko2008b} that on an equal footing accounts the laser field and the Coulomb force in the Newton's equation for electron motion in the continuum. The TCSFA applies the first-order semiclassical perturbation theory \cite{PPT1967} to account the Coulomb potential in the phase associated with every trajectory. The same first-order semiclassical perturbation theory was used in the phase of the QTMC model. In contrast to this, the SCTS and the CQSFA approaches go beyond the perturbation theory. 

The SCTS model operates with large ensembles of classical trajectories that are propagated in the continuum to find the final asymptotic momenta and bin them (and, therefore, the corresponding contributions assigned to these trajectories) in bins in momentum space. This approach is often referred to as ``shooting method" (see, e.g., Ref.~\cite{Yan2010}). Instead, the CQSFA model solves the so-called ``inverse problem", i.e., finds all the trajectories leading to a given final momentum. This allows to avoid large ensembles of trajectories and establish a better control over cusps and caustics that are inevitable in trajectory-based simulations. The price that is to be paid is that the solution of the inverse problem is a difficult task. In addition to this, the approach with the inverse problem can often be less versatile. 

In this paper we review the SCTS model, as well as two recent implementations of this model. We also discuss some of the applications of the SCTS. The SCTS model has been applied to the investigation of the intra-half-cycle interference of photoelectrons with low energies \cite{Xie2016}, to the studies of the interference patterns arising in the strong-field photoelectron holography \cite{Walt2017,Shvetsov2018a,Arbo2019}, to the analysis of the sub-cycle interference in ionization by counter-rotating two-color fields \cite{Eckart2018}, to the investigation of sideband modulation by subcycle interference in ionization by circularly polarized two-color laser fields \cite{Eckart2020}, etc. Here we focus on the applications of the SCTS to the strong-field photoelectron holography, study of the multielectron polarization effects, and the ionization of the H$_2$ molecule.  

The paper is organized as follows. In Sec.~II we review the SCTS and discuss different approaches used to implement this model numerically. In Sec.~III we discuss the further modifications of the SCTS model: the semiclassical two-step model with quantum input and the SCTS model accounting for the preexponential factor of the semiclassical propagator. In Sec.~IV we briefly review applications of the SCTS model to the strong-field photoelectron holography. The application of the SCTS to the study of the multielectron polarization effects in the ATI are discussed in Sec.~V. In Sec.~VI we review the usage of the SCTS model to describe the strong-field ionization of the H$_2$ molecule. The conclusions of this colloquia paper are given in Sec.~VII. 

\section{Semiclassical two-step model}
\subsection{Formulation of the semiclassical two-step model}
As any semiclassical model, the electron trajectory in SCTS is calculated using classical equation of motion:
\begin{equation}
\label{newton}
\frac{d^{2}\vec{r}}{dt^2}=-\vec{F}\left(t\right)-\nabla V\left(\vec{r},t\right),
\end{equation}
where $\vec{F}\left(t\right)$ is the laser field and $V\left(\vec{r},t\right)$ is the ionic potential. In order to find the trajectory from Eq.~(\ref{newton}), we need to specify the initial conditions, i.e., the initial velocity of the departing electron and the starting point. In the original version of the SCTS model it is assumed that the electron starts with zero initial velocity along the laser field $v_{0,z}=0$, but it can have a nonzero initial velocity $v_{0,\perp}$ in the perpendicular direction. We note that the application of the SFA to describe the electron motion under the potential barrier leads to a nonzero initial longitudinal velocity $v_{0,z}\neq0$. The effect of the nonzero $v_{0,z}$ will be discussed later. Let us first assume that the interaction of the ionized electron with the ion is modelled by the Coulomb potential. Then the starting point of the trajectory, i.e., the tunnel exit point, can be obtained using the separation of the static tunneling problem in parabolic coordinates. For the static field polarized along the $z$ axis we define the parabolic coordinates as $\xi=r+z$, $\eta=r-z$, and $\varphi=\atan\left(y/x\right)$ and find the tunnel exit coordinate $\eta_{e}$ from the following equation:
\begin{equation}
\label{tunex}
-\frac{\beta_{2}\left(F\right)}{2\eta}+\frac{m^2-1}{8\eta^2}-\frac{F\eta}{8}=-\frac{I_{p}\left(F\right)}{4}.
\end{equation}
Here $m$ is the magnetic quantum number of the initial state, $I_p\left(F\right)$ is the Stark-shifted ionization potential, and 
\begin{equation}
\label{beta}
\beta_{2}\left(F\right)=Z-\left(1+\left|m\right|\right)\frac{\sqrt{2I_{p}\left(F\right)}}{2}.
\end{equation}
The tunnel exit point is given by $z_{e}=-\eta_{e}/2$. In the general case, the ionization potential $I_p\left(F\right)$ in Eq.~(\ref{tunex}) is given by
\begin{equation}
\label{yp}
I_{p}\left(F\right)=I_p\left(0\right)+\left(\vec{\mu}_{N}-\vec{\mu}_{I}\right)\cdot\vec{F}+\frac{1}{2}\left(\alpha_N-\alpha_I\right)\vec{F}^{2}.
\end{equation}
Here $I_p\left(0\right)$ is the ionization potential in the absence of the field, and $\vec{\mu}_{N, I}$ and $\alpha_{N, I}$ are the dipole moments and static polarizabilities, respectively. The index $N$ refers to the neutral atom (molecule), and the index $I$ stands for its ion. We note that for atom the term linear with respect to $F$ is absent in Eq.~(\ref{yp}). The static field $F$ in Eqs.~(\ref{tunex}), (\ref{beta}), and (\ref{yp}) should be replaced by the instantaneous value of the laser field at the time of ionization $t_0$.

The instants of ionization and the initial transverse velocities are distributed in accord with the static ionization rate \cite{DeloneKrainov1991}:
\begin{equation}
\label{tunrate}
w\left(t_{0},v_{0, \perp}\right)\sim\exp\left(-\frac{2\kappa^3}{3F\left(t_0\right)}\right)\exp\left(-\frac{\kappa v_{0,\perp}^{2}}{F\left(t_0\right)}\right),
\end{equation} 
where $\kappa=\sqrt{2I_{p}}$. Following the original formulation of the SCTS model we omit the preexponential factor in Eq.~(\ref{tunrate}). For atoms it only slightly affects the shape of the electron momentum distributions.

After the laser pulse terminates an electron moves in the Coulomb field only. If the electron energy at the time $t=t_{f}$ at which the laser pulse terminates is negative $E<0$, the electron moves along the elliptical orbit, and it should be treated as captured into a Rydberg state \cite{Nubbemeyer2008,Shvetsov2009}. The corresponding process is often referred to as frustrated tunnel ionization, see, e.g., Refs.~\cite{McKenna2012,Agapi2012,Price2014,Popruzhenko2017}. It is clear that the trajectories with $E<0$ should be excluded from consideration, if we are interested in ionized electrons. 
The latter obviously correspond to the hyperbolic trajectories ($E>0$). The asymptotic momentum $\vec{k}$ of the electron is determined by its position $\vec{r}\left(t_f\right)$ and momentum $\vec{p}\left(t_f\right)$ at the time $t=t_f$:
\begin{equation}
\label{mominf}
\vec{k}=k\frac{k\left(\vec{L}\times\vec{a}\right)-\vec{a}}{1+k^2L^2},
\end{equation}
see Refs.~\cite{Shvetsov2012,Shvetsov2009}. In Eq.~(\ref{mominf}) $\vec{L}=\vec{r}\left(t_f\right)\times\vec{p}\left(t_f\right)$ and $\vec{a}=\vec{p}\left(t_f\right)\times\vec{L}-Z\vec{r}\left(t_f\right)/r\left(t_f\right)$ are the angular momentum and the Runge-Lenz vector, respectively. The magnitude of the momentum $k$ is determined by the energy conservation:
\begin{equation}
\label{enrg}
\frac{k^2}{2}=\frac{\vec{p}\left(t_f\right)}{2}-\frac{Z}{r\left(t_f\right)}.
\end{equation}

The key ingredient of the SCTS model is the expression for the phase associated with every trajectory. This phase corresponds to the phase of the matrix element of the semiclassical propagator $U_{SC}\left(t_2,t_1\right)$ between the initial state at time $t_1$ and the final state at time $t_2$ \cite{Miller1974,Walser2003,Spanner2003} (for a text-book treatment see Refs.~\cite{TannorBook,GrossmannBook}). Depending on the variables used to describe the initial and final states there exist four equivalent forms of the semiclassical propagator $U_{SC}$:
\begin{subequations}
\begin{align}
\label{Matrixelem1}
\left\langle \vec{r}_2\right|U_{\rm{SC}}\left(t_2,t_1\right)\left|\vec{r}_1\right\rangle & = \left[-\frac{\det\left(\partial^2\phi_{2}\left(\vec{r}_1,\vec{r}_2\right)/\partial \vec{r}_1 \partial \vec{r}_2\right)}{\left(2\pi i\right)^3}\right]^{1/2} \nonumber\\
& \times \exp\left[i \phi_{2}\left(\vec{r}_1,\vec{r}_2\right)\right]\, ,\\
\label{Matrixelem2}
\left\langle \vec{r}_2\right|U_{\rm{SC}}\left(t_2,t_1\right)\left|\vec{p}_1\right\rangle & = \left[-\frac{\det\left(\partial^2\phi_{2}\left(\vec{p}_1,\vec{r}_2\right)/\partial \vec{p}_1 \partial \vec{r}_2\right)}{\left(2\pi i\right)^3}\right]^{1/2} \nonumber\\
& \times \exp\left[i \phi_{2}\left(\vec{p}_1,\vec{r}_2\right)\right]\, ,\\ 
\label{Matrixelem3}
\left\langle \vec{p}_2\right|U_{\rm{SC}}\left(t_2,t_1\right)\left|\vec{r}_1\right\rangle & = \left[-\frac{\det\left(\partial^2\phi_{3}\left(\vec{r}_1,\vec{p}_2\right)/\partial \vec{r}_1 \partial \vec{p}_2\right)}{\left(2\pi i\right)^3}\right]^{1/2} \nonumber\\
& \times \exp\left[i \phi_{3}\left(\vec{r}_1,\vec{p}_2\right)\right]\, ,\\
\label{Matrixelem4}
\left\langle \vec{p}_2\right|U_{\rm{SC}}\left(t_2,t_1\right)\left|\vec{p}_1\right\rangle & = \left[-\frac{\det\left(\partial^2\phi_{4}\left(\vec{p}_1,\vec{p}_2\right)/\partial \vec{p}_1 \partial \vec{p}_2\right)}{\left(2\pi i\right)^3}\right]^{1/2} \nonumber\\
& \times \exp\left[i \phi_{4}\left(\vec{p}_1,\vec{p}_2\right)\right]\, .
\end{align}
\end{subequations}
Here $\vec{r}_1$ ($\vec{r}_2$) and $\vec{p}_1$ ($\vec{p}_2$) are the initial (final) coordinates and momenta, respectively. The phase $\phi_1$ that corresponds to the transition from the initial state to the final state, which are both described by the position, is determined by the classical action: 
\begin{equation}
\label{phi1}
\phi_{1}\left(\vec{r}_1,\vec{r}_2\right)=\int_{t_1}^{t_2}\left\{\vec{p}\left(t\right)\dot{\vec{r}}\left(t\right)-H\left[\vec{r}\left(t\right),\vec{p}\left(t\right)\right]\right\}dt,
\end{equation}
where $H\left[\vec{r}\left(t\right),\vec{p}\left(t\right)\right]$ is the classical Hamiltonian function that depends on the canonical coordinates $\vec{r}\left(t\right)$ and momenta $\vec{p}\left(t\right)$. The other three phases $\phi_2$, $\phi_3$, and $\phi_4$ are related to $\phi_1$ by the canonical transformations: 
\begin{subequations}
\begin{align}
\label{phi2}
\phi_2\left(\vec{p}_1,\vec{r}_2\right)&=\phi_{1}\left(\vec{r}_1,\vec{r}_2\right)+\vec{p}_{1}\cdot\vec{r}_{1}\, ,\\
\label{phi3}
\phi_3\left(\vec{r}_1,\vec{p}_2\right)&=\phi_{1}\left(\vec{r}_1,\vec{r}_2\right)-\vec{p}_{2}\cdot\vec{r}_{2}\, ,\\
\label{phi4}
\phi_4\left(\vec{p}_1,\vec{p}_2\right)&=\phi_{1}\left(\vec{r}_1,\vec{r}_2\right)+\vec{p}_{1}\cdot\vec{r}_{1}-\vec{p}_{2}\cdot\vec{r}_{2}\, ,
\end{align}
\end{subequations}
Then the question arises: Which of these phases should be chosen for description of the strong-field ionization process? On the assumption that for a given ionization time the starting-point of the electron trajectory is localized in space [see Eq.~(\ref{tunex})] and the final state is characterized by the asymptotic momentum $\vec{k}$, the phase $\phi_3$ is used in the SCTS model. Indeed, the strong-field ionization can be viewed as a half-scattering process of an electron that is initially localized near the atom (molecule) and detected with the final momentum $\vec{k}$. We note that if the initial longitudinal velocity is equal to zero, the initial electron momentum $\vec{p}_1$ is orthogonal to the initial position vector $\vec{r}_1$ (i.e., $\vec{p}_{1}\cdot\vec{r}_{1}=0$), and therefore, the phases $\phi_3$ and $\phi_1$ coincide with each other. For nonzero $v_{0,z}$ the term $\vec{p}_{1}\cdot\vec{r}_{1}$ is to be accounted in the phase. However, in most cases this term almost does not affect the resulting electron momentum distributions.

As the result, the phase corresponding to a given trajectory in the SCTS model is given by:
\begin{align}
\label{phas_scts}
& \Phi^{SCTS}\left(t_{0},\vec v_0\right)= -\vec v_0\cdot\vec r(t_0) + I_{p}t_{0} \nonumber\\
& -\int_{t_0}^\infty dt\, \left\lbrace\dot{\vec p}(t)\cdot\vec{r}(t)+H[\vec{r}\left(t\right),\vec{p}(t)]\right\rbrace,  
\end{align}
where it is assumed that the trajectory has also the initial phase $\exp\left(iI_{p}t_0\right)$ that describes the time evolution of the ground state. The expression (\ref{phas_scts}) can be also written as follows:
\begin{align}
\label{phas_sim}
& \Phi^{SCTS}\left(t_{0},\vec v_0\right)= - \vec v_0\cdot\vec r(t_0) + I_{p}t_{0} \nonumber \\
& - \int_{t_0}^\infty dt\, \left\lbrace\frac{p^2(t)}{2}+V[\vec{r}(t)]-\vec r(t)\cdot\vec\nabla V[\vec{r}(t)]\right\rbrace.
\end{align}
To arrive at the expression (\ref{phas_sim}), we use the explicit form of the Hamiltonian for an arbitrary effective potential 
\begin{equation}
\label{hamilt}
H\left[\vec{r}\left(t\right),\vec{p}\left(t\right)\right]=\frac{\vec{p}^2\left(t\right)}{2}+\vec{F}\left(t\right)\cdot\vec{r}\left(t\right)+V\left(\vec{r}\right)
\end{equation}
and employed Newton’s equation of motion (\ref{newton}). This formula is applicable for any single-active-electron potential used to describe the multielectron system (atom or molecule), including pseudopotentials (see, e.g., Ref.~\cite{Tong2015} and references therein). For the specific case of the Coulomb potential, the phase (\ref{phas_sim}) reads as
\begin{eqnarray}
\label{phas_coul}
\Phi^{STCS}\left(t_{0},\vec v_0\right)&=& - \vec v_0\cdot\vec r(t_0) + I_{p}t_{0}\nonumber \\
&-&\int_{t_0}^\infty dt\, \left\lbrace\frac{p^2(t)}{2}+\frac{2Z}{r\left(t\right)}\right\rbrace.
\end{eqnarray}
This formula should be compared with the phase used in the QTMC model:
\begin{eqnarray}
\label{phas_qtmc}
\Phi^{QTMC}\left(t_{0},\vec v_0\right)&=& - \vec v_0\cdot\vec r(t_0) + I_{p}t_{0}\nonumber \\
&-&\int_{t_0}^\infty dt\, \left\lbrace\frac{p^2(t)}{2}+\frac{Z}{r\left(t\right)}\right\rbrace.
\end{eqnarray}
It is seen that the QTMC phase can be obtained from Eq.~(\ref{phas_coul}) by neglecting the term $\vec r(t)\cdot\vec\nabla V[\vec{r}(t)]$ in the integrand. This term leads to the double weight of the Coulomb term in the SCTS compared to the QTMC. Therefore, the QTMC phase can be considered as an approximation to the SCTS one. The double weight of the Coulomb contribution leads to a better agreement with the TDSE results \cite{Shvetsov2016}.

The SCTS phase (\ref{phas_coul}) is divergent at $t\to\infty$, and therefore, it is to be regularized. The regularization (see Ref.~\cite{Shvetsov2016}) can be accomplished by decomposing the SCTS phase as
\begin{align}
\label{phase_decom}
&\Phi^{SCTS}\left(t_{0},\vec v_0\right)=-\vec v_0\cdot\vec r(t_0) + I_{p}t_{0}\nonumber\\
&-\int_{t_0}^{t_f}\left\{\frac{p^2\left(t\right)}{2}-\frac{2Z}{r\left(t\right)}\right\}-\int_{t_f}^{\infty}dt\left\{E-\frac{Z}{r\left(t\right)}\right\}
\end{align}
and separating the time-independent part of the integrand in the term $\int_{t_f}^{\infty}dt\left\{E-Z/r\left(t\right)\right\}$. Although this time-independent part leads to the contribution \\ $\lim\limits_{t\to\infty}E\left(t-t_f\right)$ that diverges linearly when $t\to\infty$, it does not produce a phase difference for electron trajectories ending up in the same bin. Indeed, the final momenta of such trajectories (and, therefore, their energies) should be considered as equal. Using the solution of the Kepler problem (see, e.g., Ref.~\cite{Dau1}) we calculate the divergent integral 
\begin{equation}
\label{div_con}
\Phi_{f}^{C}\left(t_f\right)=Z\int_{t_f}^{\infty}\frac{dt}{r\left(t\right)}
\end{equation}
analytically: $\Phi_{f}^C\left(\tau_f\right)=Z\sqrt{b}\left[\xi\left(\infty \right)-\xi\left(\tau_f\right)\right]$. The parameter $\xi$ is used to parametrize the time $t$ and the distance $r$ from the Coulomb center:
\begin{eqnarray}
\label{Coul_prob}
& & r\left(t\right)=b\left(g \cosh\xi-1\right),\nonumber\\
& & t=\sqrt{b^3}\left(g\sinh \xi-\xi \right)+C
\end{eqnarray}
Here, in turn, $b=1/\left(2E\right)$ and $g=\sqrt{1+2EL^2}$. The constant $C$ in Eq.~(\ref{Coul_prob}) is to be found using the initial conditions, i.e, $\vec{r}\left(t_f\right)$ and $\vec{p}\left(t_f\right)$. It is easy to verify that 
\begin{equation}
\xi\left(t_f\to\infty\right)=\ln\left(\frac{2t}{g\sqrt{b^3}}\right),
\end{equation}
see Ref.~\cite{Shvetsov2016}. Therefore, for trajectories arriving at the same bin, we can discriminate between the common divergent part $\ln\left(2t/\sqrt{b^3}\right)$ and the finite contributions determined by $-\ln\left(g\right)$. We note that the latter depends not only on the energy, but also on the angular momentum $L$, and thus is different for different trajectories interfering in a given bin. Since 
\begin{equation}
\label{up_lim}
\xi\left(t_f\right)=\arcsinh\left\{\frac{\vec{r}\left(t_f\right)\cdot\vec{p}\left(t_f\right)}{g\sqrt{b}}\right\},
\end{equation}
we obtain the following contribution to the phase accumulated in the time interval $\left[t_f,\infty\right]$ due to the Coulomb potential (see Ref.~\cite{Shvetsov2016}):
\begin{equation}
\label{post_pulse}
\tilde{\Phi}_{f}^{C}\left(t_f\right)=-Z\sqrt{b}\left[ \ln g+\arcsinh\left\{\frac{\vec{r}(t_f)\cdot\vec{p}(t_f)}{g\sqrt{b}}\right\}\right].
\end{equation}
This asymptotic correction of the phase which we call post-pulse phase is missing in the QTMC model. 

\subsection{Implementation of the semiclassical two-step model}

The expression for the phase can be conveniently treated as an additional equation in the system of the first order ordinary differential equations for electron coordinates and velocity components following from (\ref{newton}). This system can be solved using the fourth-order Runge-Kutta method with adaptive step size \cite{NumRep}. The ability of the numerical method to change the integration step is particularly important at small distances from the Coulomb center.

It is clear that the convergence of the results must be controlled with respect to both the size of the bin in the momentum space and the number of trajectories. It is particularly convenient to control convergence by using the energy spectra. In contrast to the three-dimensional (3D) differential momentum distributions or their two-dimensional (2D) cuts, the spectra are functions of only one variable. They can be easily compared to each other in, e.g., logarithmic scale. 

Already the first practical application of the SCTS model has shown that a large number of trajectories is needed for convergence (see Ref.~\cite{Shvetsov2016} for details). Typically, for the same laser parameters a thousand times more trajectories are needed for the simulations with the phase compared to a semiclassical model disregarding the interference effect. This can be expected taking into account the fine interference details of electron momentum distributions generated in ionization by strong laser pulses. For this reason, it is important to consider optimization of the codes implementing the SCTS model. The most obvious way to speed up the SCTS calculations is to use parallelization. Indeed, any trajectory-based simulation can be very easily and efficiently implemented on a computer cluster by paralleling the loop over the number of trajectories.

Another approach consists in an efficient sampling of the initial conditions, i.e., times of ionization $t_0^{j}$ and initial velocities $v_0^{j}$, where index $j$ enumerates the trajectories of an ensemble. In a standard trajectory-based approach the initial conditions are chosen either randomly or from a certain uniform grid. Neglecting interference effect the ionization probability $R\left(\vec{k}\right)$ for the final momentum $\vec{k}$ that corresponds to the bin $\left[k_{i}, k_{i}+\Delta k_{i}\right]$ $\left(i=x,y,z\right)$ is calculated as
\begin{equation} 
\label{stad_noint}
R\left(\vec{k}\right)=\sum_{j=1}^{n_p}w\left(t_0^{j},v_0^{j}\right),
\end{equation}
while the similar formula for the SCTS model reads as
\begin{eqnarray} 
\label{stad_int}
& & R\left(\vec{k}\right)=\nonumber \\
& & \sum_{j=1}^{n_p}\left|\sqrt{w\left(t_0^{j},v_0^{j}\right)}\exp\left[i\Phi^{SCTS}\left(t_0^{j},v_0^{j}\right)\right]\right|^2.
\end{eqnarray}
The sums in Eqs.~(\ref{stad_noint}) and (\ref{stad_int}) are calculated over all $n_p$ trajectories arriving at the given bin. However, the approach sketched here is not the only possible one. Importance sampling widely used in Monte-Carlo integration (see, e.g., Ref.~\cite{NumRep}) can be used to implement the SCTS model.

We turn first to the semiclassical simulations disregarding interference. In the important sampling approach the weights (importance) of classical trajectories are accounted already at the sampling stage. More specifically, the sets of initial conditions $\left(t_0^{j},v_0^{j}\right)$ are distributed in accord with the tunneling rate $w\left(t_{0}^{j},v_{0}^{j}\right)$ and the ionization probability $R\left(\vec{k}\right)$ is given by a number of trajectories reaching the bin corresponding to the final momentum $\vec{k}$. It is easy to see that the ionization probability in the SCTS model based on the importance sampling reads as
\begin{equation}
R\left(\vec{k}\right)=\sum_{j=1}^{n_p}\left|\exp\left[i\Phi^{SCTS}\left(t_0^{j},v_0^{j}\right)\right]\right|^2.
\end{equation}
with the initial conditions distributed in accord to the square root of the ionization probability. In many situations the important sampling technique provides faster convergence compared to the standard approach of \\ Eqs.~(\ref{stad_noint})-(\ref{stad_int}). Its performance, however, depends on the laser-atom parameters and the specific part of photoelectron momentum distribution under study.

\subsection{Benchmark case: Ionization of the H atom}

The SCTS model was compared with the QTMC approach and direct numerical solution of the TDSE for ionization of the hydrogen atom (see Ref.~\cite{Shvetsov2016}). The 2D electron momentum distributions calculated in accord to the all three approaches are shown in Fig.~2~(a)-(c). The simulations are done for ionization by a few-cycle laser pulse linearly polarized along the $z$-axis and defined through the vector-potential:
\begin{equation}
\label{vecpot}
\vec{A}\left(t\right)=\left(-1\right)^{n}\frac{F_0}{\omega}\sin^2\left(\frac{\omega t}{2n}\right)\sin\left(\omega t+\varphi\right)\vec{e}_{z}.
\end{equation}
Here $n$ is the number of optical cycle within the pulse, and $\vec{e}_z$ is the unit vector in the polarization direction. The pulse (\ref{vecpot}) is present between $t=0$ and $t=t_f=\left(2\pi/\omega\right) \cdot n$. The factor $\left(-1\right)^n$ in (\ref{vecpot}) ensures that the field has its maximum at the center of the pulse ($\omega t=\pi n$) for even and odd $n$. The laser field is to be calculated from Eq.~(\ref{vecpot}) as $\vec{F}\left(t\right)=-d\vec{A}/dt$.

\begin{figure}[h]
\begin{center}
\includegraphics[width=0.48\textwidth,trim={0 2cm 3cm 0cm}]{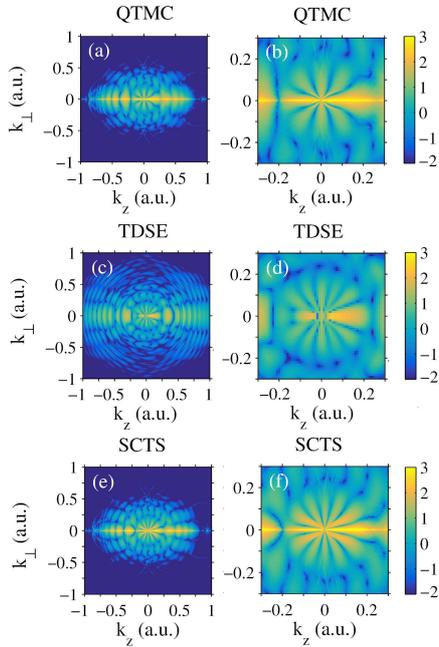} 
\end{center}
\caption{Two-dimensional electron momentum distributions for ionization of the H atom by a laser pulse with a duration of $n=8$ cycles, peak intensity of $0.9\times10^{14}$ W/cm$^2$, and wavelength of 800 nm calculated from the QTMC model [(a),(b)], numerical solution of the TDSE [(c),(d)], and the SCTS [(e),(f)]. Panels (b), (d), and (f) display the magnifications for $\left|k_z\right|$,$\left|k_{\perp}\right|<0.3$~a.u. of the distributions shown in (a), (c), and (d), respectively. The laser pulse is linearly polarized along the $z$ axis. The distributions are normalized to the total ionization yield. A logarithmic color scape in arbitrary units is used.}
\label{fig1}
\end{figure}

It is seen that the most important features of the TDSE result are reproduced by the semiclassical models [see Figs.~\ref{fig1}~(a),(c),and (f)]. Indeed, the electron momentum distributions are stretched along the $z$-axis and show clear ATI rings as well as the pronounced interference structure in their low-energy parts. The width of the momentum distributions along the polarization direction is obviously underestimated by both semiclassical models. This is due to the initial condition $v_{0,z}=0$ (see Ref.~\cite{Shvetsov2016} for details). 

\begin{figure}[h]
\begin{center}
\includegraphics[width=0.35\textwidth]{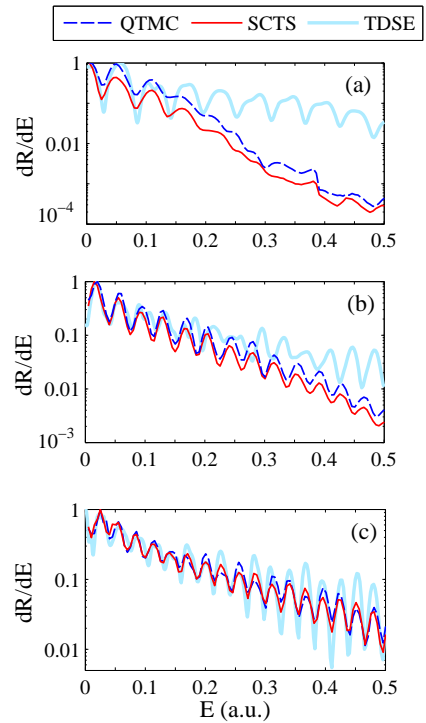} 
\end{center}
\caption{Photoelectron energy spectra for ionization of the H atom by a laser pulse with a duration of $n=8$ cycles and peak intensity of $0.9\times10^{14}$ W/cm$^2$ calculated using the TDSE (thick light blue curve), the QTMC (dashed blue curve) and the SCTS (solid red curve). Panels (a), (b), and (c) correspond to the wavelengths of 800 nm, 1200 nm, and 1600 nm, respectively. The spectra are normalized to the peak value.}
\label{fig2}
\end{figure}

\begin{figure}[h]
\begin{center}
\includegraphics[width=0.4\textwidth]{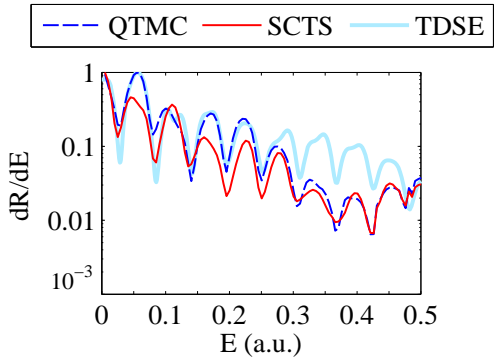} 
\end{center}
\caption{Photoelectron energy spectra calculated from the TDSE (thick light blue curve), the QTMC model (dashed blue curve) and the SCTS (solid red curve). A nonzero initial parallel velocity predicted by the SFA is used in both the QTMC and SCTS simulations. The pulse parameters are as in Fig~\ref{fig2}(a).}
\label{fig3}
\end{figure}

However, a closer examination of the low-energy parts of the distributions reveals remarkable deviations Indeed, for $\left|k\right|<0.3$~a.u. the photoelectron momentum distributions demonstrate pronounced fanlike interference structures, see Fig.~\ref{fig1}~(b), (d), and (e). These structures are similar to the ones of Ramsauer-Townsend diffraction oscillations, see Refs.~\cite{Arbo2006,Moshammer2009,Arbo20062,Arbo2008}. It is seen that the SCTS model reproduces the interference pattern of the TDSE, whereas the QTMC model predicts fewer nodal lines. This fact was attributed to the underestimate of the Coulomb potential in the expression for the phase used in the QTMC model. 
The comparison of the photoelectron energy spectra $dR/dE$ shows that the QTMC and the SCTS qualitatively reproduce the ATI peaks, see Figs.~\ref{fig2}~(a)-(c). However, both semiclassical approaches can quantitatively reproduce the amplitude of interference oscillations only for a few low-order peaks. This is related to the fact that due to the initial conditions [Eq.~(\ref{tunrate})] used in both semiclassical models too few trajectories with large initial momenta in the polarization direction are launched. This also explains why the semiclassical energy spectra fall off too rapidly with the increase of energy. In order to test this hypothesis, the initial longitudinal velocity for every ionization time is set to the value predicted by the SFA, see, e.g. Ref.~\cite{Yan2010}. This change in initial conditions leads to a better agreement between the SCTS model and the TDSE, see Fig.~\ref{fig3} and Ref.~\cite{Shvetsov2016}. Therefore, the main reason of deviations of the SCTS results from the TDSE solutions is not the semiclassical treating of the electron motion in the continuum, but the fact that the SCTS model does not describe the tunneling step accurately enough. 

\section{Modifications of semiclassical two-step model}

Substantial efforts have been recently made to modify the SCTS model. These modifications are aimed at providing not only a qualitative, but also a quantitative agreement with the TDSE. To achieve this goal, it is necessary to overcome the deficiencies of the SCTS model (as well as of any other semiclassical model) in description of the ionization step. The simplest way is to use the SFA formulas to distribute the initial conditions of classical trajectories. This approach dates back to the studies of Refs.~\cite{Yudin2001,Bondar2009}. It is used in the various semiclassical models (see, e.g., Refs.~\cite{Yan2010,Boge2013,Hofmann2014,Geng2014}), as well as in the implementations of the SCTS model developed in Refs.~\cite{Brennecke2020,Eckart2020}. We note, however, that the validity of the SFA formulas used as initial conditions for classical trajectories requires a systematical study. To the best of our knowledge, such a study has not been accomplished so far. Here we discuss two modifications of the SCTS model: The semiclassical two-stem model with quantum input (SCTSQI) \cite{Shvetsov2019b} and the SCTS model with the prefactor \cite{Brennecke2020}. 

\subsection{Semiclassical two-step model with quantum input}

The SCTSQI model combines the SCTS with initial conditions obtained from the solution of the TDSE. Such a combination leads to a novel quantum-classical approach. The SCTSQI model is formulated for ionization of a one-dimensional (1D) model atom. Therefore, before reviewing the SCTSQI, we briefly discuss the solution of the 1D TDSE, as well as the application of the SCTS model in 1D case.  

For the 1D model, the TDSE in the velocity gauge is given by
\begin{eqnarray}
\label{tdse}
& & i\frac{\partial}{\partial t}\Psi\left(x,t\right)\nonumber\\
&=& \left\{\frac{1}{2}\left(-i\frac{\partial}{\partial x}+A_{x}\left(t\right)\right)^2+V\left(x\right)\right\}\Psi\left(x,t\right),
\end{eqnarray}
where $\Psi\left(x,t\right)$ is the wave function in coordinate space. The 1D soft-core Coulomb potential 
\begin{equation}
\label{soft_pot}
V=-\frac{1}{\sqrt{x^2+a^2}}
\end{equation}
with $a=1.0$ (see Ref.~\cite{Javanainen1988}) is used in Ref.~\cite{Shvetsov2019b}. The corresponding time-independent Schrodinger equation reads as:
\begin{equation}
\label{tise}
\left\{-\frac{1}{2}\frac{d^2}{dx^2}+V\left(x\right)\right\}\Psi\left(x\right)=E\Psi\left(x\right).
\end{equation}
Equation (\ref{tise}) can be easily solved on a grid using, e.g., the well-known three-step formula for approximation of the second derivative and subsequent diagonalization. In Ref.~\cite{Shvetsov2019b} the TDSE (\ref{tdse}) is solved using slit-operator method \cite{Feit1982}. 
In the regions $x_b\le\left|x\right|\le x_{\text{max}}$ the wave function is multiplied by a mask
\begin{equation}
\label{mask}
M\left(x\right)=\cos^{1/6}\left[\frac{\pi\left(\left|x\right|-x_{b}\right)}{2\left(x_{\text{max}}-x_{b}\right)}\right],
\end{equation}
where $x=\pm x_b$ correspond to the internal boundaries of the absorbing regions, and $x_{\text{max}}$ is the size of the computational box. The mask prevents unphysical reflections of the propagating wave function from the grid boundary and allows to calculate the electron momentum distributions using the mask method \cite{Tong2006}.

In the 1D case the Newton's equation for an electron moving in the laser field and the field of the potential (\ref{soft_pot}) reads as
\begin{equation}
\frac{d^{2}x}{dt^2}=-{F}_{x}\left(t\right)-\frac{x}{\left(x^2+a^2\right)^{3/2}}.
\label{newton_1d}
\end{equation}
The corresponding SCTS phase is given by (see Ref.~\cite{Shvetsov2019b}):
\begin{eqnarray}
& & \Phi^{SCTS}\left(t_{0},\vec{v}_{0}\right)=I_{p}t_{0}\nonumber\\
&-& \int_{t_0}^\infty dt \left\{\frac{v_{x}^{2}\left(t\right)}{2}-\frac{x^2}{\left(x^2+a^2\right)^{3/2}}-\frac{1}{\sqrt{x^2+a^2}}\right\}.
\end{eqnarray}
We note that the ionization rate (\ref{tunrate}) in the 1D case is to be replaced by
\begin{equation}
w\left(t_0\right)\sim\exp\left(-\frac{2\left(2\left|E_{0}\right|\right)^{3/2}}{3F\left(t_0\right)}\right),
\end{equation}
where $E_0=-0.6698$~a.u. is the ground-state energy in the potential (\ref{soft_pot}). Equation (\ref{newton_1d}) is to be numerically integrated up to the end of the laser pulse at $t=t_f$. The asymptotic momentum of the photoelectron can be found from $x\left(t_f\right)$ and $p_x\left(t_f\right)$ using the energy conservation law. We note that after the end of the pulse the unbound electron cannot change its direction of motion, and, therefore, $k_x$ has the same sign as that of $p_{x}\left(t_f\right)$.  

In order to correctly apply the SCTS model in the 1D case, the post-pulse phase is to be calculated. This calculation can be performes as follows (see Ref.~\cite{Shvetsov2019b}). At first, we decompose the phase as:
\begin{eqnarray}
\label{decomp}
& & \Phi^{SCTS}\left(t_{0},\vec{v}_{0}\right)=I_{p}t_{0}\nonumber\\
&-& \int_{t_0}^{t_f} dt \left\{\frac{v_{x}^{2}\left(t\right)}{2}-\frac{x^2}{\left(x^2+a^2\right)^{3/2}}-\frac{1}{\sqrt{x^2+a^2}}\right\}\nonumber\\
&+&\Phi_{f}^{V},
\end{eqnarray}
As in the 3D case, we separate the post-pulse phase into parts with time-dependent and time-independent integrands and disregard the linearly divergent contribution from the first part. As the result, the post-pulse phase is determined by:
\begin{equation}
\label{postpulse2}
\tilde{\Phi}_{f}^{V}=\int_{t_f}^{\infty}\frac{x^2\left(t\right)}{\left[x^2\left(t\right)+a^2\right]^{3/2}}dt.
\end{equation}
The divergent part of this integral can be efficiently isolated. Indeed, Eq.~(\ref{postpulse2}) can be equivalently rewritten as follows:
\begin{eqnarray}
\label{regul}
& & \tilde{\Phi}_{f}^{V}=\int_{t_f}^{\infty}\left[\frac{x^{2}}{\left(x^2+a^2\right)^{3/2}}-\frac{2Et^{2}}{\left(2Et^2+a^2\right)^{3/2}}\right]dt\nonumber\\
&+&\int_{t_f}^{\infty}\frac{2Et^{2}}{\left(2Et^2+a^2\right)^{3/2}}dt.
\end{eqnarray}
Since the second divergent term in Eq.~(\ref{regul}) depends on the electron energy $E$ and the parameter $a$, it is the same for every trajectory that arrives at a given bin $[k_x-\Delta k_x, k_x+\Delta k_x]$. Therefore, it does not affect the resulting interference pattern and can be omitted \cite{Shvetsov2019b}. The post-pulse phase is determined by the first term in Eq.~(\ref{regul}). This converging integral is easily calculated numerically. It depends on the position $x\left(t_f\right)$ and velocity $p_x\left(t_f\right)$ at the end of the laser pulse what suggests an efficient way to calculate it by interpolation \cite{Shvetsov2019b}. 

This is not a simple task to unify the direct solution of the TDSE and the trajectory-based approach in one single model. The main problem of such combination has a fundamental origin. Indeed, both the starting point and the initial velocity are needed to uniquely determine the classical trajectory.
On the other hand, the Heisenberg's uncertainty principle imposes a limit to the precision with which position and momentum (as other canonically conjugated variables) can be simultaneously known. The application of quasiprobability distribution allows to extract the information from the wave function about both the coordinate and momentum.

The most widely-known examples of the quasiprobability distributions are the Wigner function and Husimi distribution \cite{Husimi} (see Ref.~\cite{Ballentine} for a textbook treatment). The latter can be obtained by smoothing of the Wigner function with a Gaussian weight. The Gabor transformation \cite{Gabor} was used in Ref.~\cite{Shvetsov2019b}. The Gabor transformation is presently widely used in studies of the ATI (see, e.g., Ref.~\cite{Shu2016}) and, especially, the HHG (see, e.g., Refs.~\cite{Chirila2010,Bandrauk2012,Wu2013}). The Gabor transform of the wave function $\tilde{\Psi}\left(x,t\right)$ near the point $x_0$ is given by:
\begin{eqnarray}
\label{Gabor}
& & G\left(x_0,p_x,t\right)=\frac{1}{\sqrt{2\pi}}\int_{-\infty}^{\infty}\tilde{\Psi}\left(x^{\prime},t\right)\exp\left[-\frac{\left(x^{\prime}-x_{0}\right)^2}{2\delta_{0}^{2}}\right]\nonumber\\
&\times&\exp\left(-ip_xx^{\prime}\right)dx^{\prime},
\end{eqnarray}
where $\delta_0$ is the width of the Gaussian window. 
The square modulus $\left|G\left(x_0,p_x,t\right)\right|^2$ corresponds to the momentum distribution of the particle in the vicinity of $x=x_0$ at time $t$ and is just the Husimi distribution \cite{Husimi}. The Hisimi distribution is a positive semidefinite function, which helps to interpret it as a quasiprobability distribution. 

\begin{figure}[h]
\begin{center}
\includegraphics[width=0.45\textwidth]{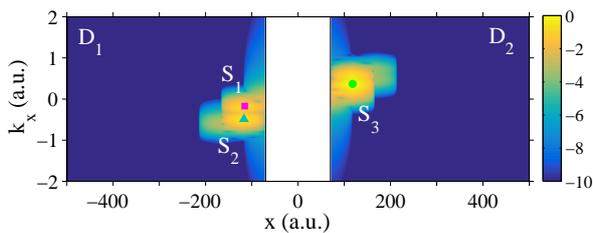} 
\end{center}
\caption{The Husimi quasiprobability distribution $\left|G\left(x,p_x,t\right)\right|^2$ at $t=3t_{f}/2$ calculated for ionization of 1D model atom by a laser pulse with a duration of $n=4$ cycles, wavelength of 800 nm, and peak intensity of $2.0\times10^{14}$~W/cm$^2$. The distribution is calculated in the phase space domains $D_1$ and $D_2$ (see text). The points $S_1$, $S_2$, and $S_3$ depicted by a green circle, magenta square, and cyan rectangle, respectively, show the three main maxima of the Husimi distribution. A logarithmic color scale is used.}
\label{fig4}
\end{figure}

The SCTSQI model employs the solution of the TDSE in the length gauge:
\begin{eqnarray}
& &i\frac{\partial}{\partial t}\Psi\left(x,t\right)\nonumber\\
&=&\left\{-\frac{1}{2}\frac{\partial^2}{\partial x^2}+V\left(x\right)+F_{x}\left(t\right)x\right\}\Psi\left(x,t\right).
\label{tdse_len}
\end{eqnarray}
Two additional spatial grids containing $N$ points are introduced the absorbing regions $\left|x\right| \ge\ x_{b}$:
\begin{equation}
\label{grids}
x_{0,\pm}^{j}=\mp\left(x_{b}+\Delta x \cdot j\right),
\end{equation}
Here $j=0,...,N$ and $\Delta x=\left(x_{max}-x_{b}\right)/N$. In the SCTSQI the Gabor transforms of the absorbed part of the wave function $\tilde{\Psi}\left(x,t\right)=\left[1-M\left(x\right)\right]\Psi\left(x,t\right)$ are calculated at every time at the points $x_{0,-}^{j}$ and $x_{0,-}^{j}$ of the grids (\ref{grids}). The value of the Gabor transformation at an arbitrary point belonging to $D_1$ or $D_2$ can be obtained by interpolation (see Ref.~\cite{Shvetsov2019b} for a details of the implementation of the SCTSQI model). Hence, at every time $t$ the Gabor transform $G\left(x,p_x,t\right)$ is known on the grids in the phase-space domains $D_{1}=\left[-x_{\text{max}},-x_{b}\right]\times\left[-p_{x,\text{max}}, p_{x,\text{max}}\right]$ and $D_{2}=\left[x_{b},x_{\text{max}}\right]\times\left[-p_{x,\text{max}}, p_{x,\text{max}}\right]$. An example of the Husimi distribution obtained in the domains $D_1$ and $D_2$ at $t=3t_{f}/2$ is shown in Fig.~\ref{fig4}.  It should be stressed that the size of the computational box $x_{max}$ used in the SCTSQI can be much smaller than the one required to obtain accurate momentum distributions by using the mask method. 

At every time $t_0$ an ensemble of $n_p$ classical trajectories with random initial positions $x_0^{j}$ and momenta $p_{x,0}^{j}$ $\left(j=1,...,n_p\right)$ is launched in the SCTSQI model. Every trajectory of the ensemble is assigned with the amplitude $G\left(t_0,x_{0}^{j},p_{x,0}^{j}\right)$ and the SCTSQI phase
\begin{eqnarray}
\label{phase_sctsqi}
& &\Phi^{SCTSQI}_{0}\left(t_{0},x_{0}^{j},p_{x,0}^{j}\right)\nonumber\\
&=&-\int_{t_0}^{\infty} dt \left\{\frac{v_{x}^{2}\left(t\right)}{2}-\frac{x^2}{\left(x^2+a^2\right)^{3/2}}-\frac{1}{\sqrt{x^2+a^2}}\right\}.
\end{eqnarray}
This phase coincides with the phase of the semiclassical propagator describing a transition from an initial state characterized by the momentum to a final state, which is also described by the momentum value. The ionization probability $R\left(k_{x}\right)$ is calculated as:
\begin{eqnarray}
\label{sctsqi}
& &R\left(k_x\right)=\left|\sum_{m=1}^{N_{T}}\sum_{j=1}^{n_{k_x}}G\left(t_{0}^{m},x_{0}^{j},p_{x,0}^{j}\right)\right. \nonumber\\ 
&\times& \left.\exp\left[i\Phi^{SCTSQI}\left(t_{0}^{m},x_{0}^{j},p_{x,0}^{j}\right)\right]\vphantom{\frac{1}{2}}\right|^{2} ,
\end{eqnarray} 
where $N_T$ is the number of steps that is used in the TDSE propagation and $n_k$ is the number of trajectories arriving at the same bin centered at $k_x$. It is important to stress that $G\left(t_{0}^{m},x_{0}^{j},p_{x,0}^{j}\right)$ is a complex function having both modulus and the phase. 

\begin{figure}[h]
\begin{center}
\includegraphics[width=0.45\textwidth]{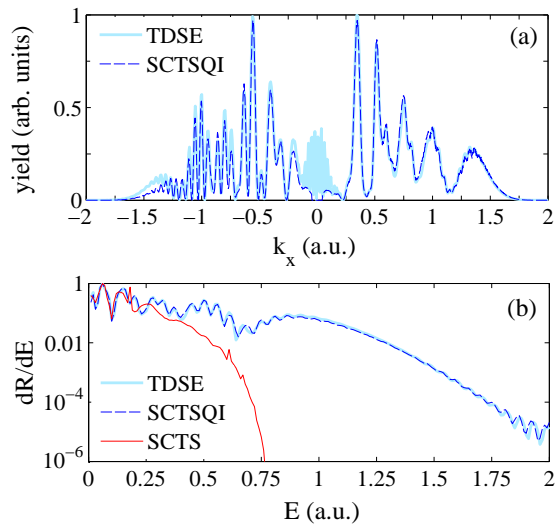} 
\end{center}
\caption{(a) The photoelectron momentum distributions for ionization of a 1D atom by a laser pulse with a duration of $n=4$ cycles, wavelength
of 800 nm, and peak intensity of $2.0\times10^{14}$ W/cm$^2$ calculated from the solution of the TDSE (thick light blue curve) and the SCTSQI model (dashed green curve). (b) Electron energy spectra obtained from the TDSE (thick light blue curve), SCTSQI (dashed green curve), and the SCTS (red curve). The distributions and spectra are normalized to the peak values.}
\label{fig5}
\end{figure}

The SCTSQI model was tested by comparing its predictions with the numerical solution of the TDSE and the SCTS model, see Figs.~\ref{fig5}~(a) and (b). It is seen that the SCTSQI provides not only qualitative, but also quantitative agreement with the TDSE result. This is true for both the width of the electron momentum distributions and the positions of the interference maxima and minima. The small discrepancy in the heights of some interference peaks [see in Fig.~\ref{fig5}~(a)] is attributed to the fact that the SCTSQI model does not account for the preexponential factor of the semiclassical matrix element. We note that as in the 3D case (see Sec.~2.3) the 1D SCTS model shows only a qualitative agreement with the fully quantum results, see Fig.~\ref{fig5} (b). Specifically, the SCTS model underestimates the width of the momentum distributions. The electron energy spectra calculated within the SCTSQI model and from the solution of the TDSE are in almost perfect agreement. Simultaneously, the spectrum calculated using the SCTS model falls off too rapidly with the increase of the energy. This is caused by the underestimation of the width of the electron momentum distributions in the SCTS model.
It was shown that the phase of the Gabor transform is very important in the SCTSQI \cite{Shvetsov2019b}. Without this phase the SCTSQI model does not provide even a qualitative agreement with the TDSE result. This could be expected, since the amplitude $G\left(t,x,p_x\right)$ contains all the information about the quantum dynamics of the absorbed part of the wave function before it was transformed in an ensemble of trajectories. In a way the term $I_pt_0$ in the phase (\ref{phas_scts}) of the SCTS model plays the same role as the phase of $G\left(t,x,p_x\right)$ in the SCTQI approach.

As any semiclassical approach, the SCTSQI model can visualize the physical mechanism responsible for the strong-field process under study using classical trajectories (see Ref.~\cite{Shvetsov2019b} for details). Since the initial conditions in the SCTSQI model are determined from the direct solution of the TDSE, we expect that this model will be able to provide more accurate trajectory-based pictures of strong-field phenomena compared to the standard semiclassical approaches. This advantage of the SCTSQI model should be used in studies of complicated strong-field processes. In addition to this, after some modification the SCTSQI model can be applied to studies of the rescattering-induced phenomena, especially the high-order ATI and the HHG. The ways of this modification are suggested in Ref.~\cite{Shvetsov2019b}. Finally, the extension of the SCTSQI model to the three-dimensional (3D) case is straightforward and developments in this direction are on the way. Most importantly, the model solves the non-trivial problem how to choose initial conditions for classical trajectories. In the SCTSQI model these initial conditions are determined by the exact quantum dynamics.

\subsection{SCTS model with preexponential factor}

An efficient modification and extension of the SCTS model was proposed recently in Ref.~\cite{Brennecke2020}. This study for the first time investigates systematically the influence of the  preexponential factor of the semiclassical matrix element (\ref{Matrixelem3}) (see Refs.~\cite{Levit1977,Levit1978}) that was not explicitly considered in all other versions of the SCTS. The modulus of this prefactor that corresponds to the mapping from initial conditions to the final momentum components influences the weights of the classical trajectories. Its phase known as the Maslov phase can be identified as a case of Gouy's phase anomaly and modifies the interference structures \cite{Brennecke2020}. In addition, the authors propose a novel way of solving the so-called inverse problem based on a clustering algorithm. 

Since the SCTS implementation of Ref.~\cite{Brennecke2020} employs the SFA and the saddle-point approximation to calculate the ionization weight of the classical trajectories and their initial positions, the ionization time $t_0$ for each initial electron momentum $\vec{k}'$ is determined by the real part of the corresponding saddle-point time $t_s=t_0+it_1$. The saddle point $t_s$ satisfies the equation:
\begin{equation}
\frac{1}{2}\left[\vec{k}'+\vec{A}\left(t_s\right)\right]^2+I_{p}=0.
\end{equation}
The ionization probability is calculated as:
\begin{eqnarray}
\label{scts_new}
& & R\left(\vec{k}\right)=\nonumber\\
& &\left|\sum\frac{DC_{\textrm{Coul}}}{\sqrt{|J\left(t\rightarrow\infty\right)|}}\exp\left[i\left(S^{0}_{\downarrow}+S_{\rightarrow}-\frac{\nu\pi}{2}\right)\right]\right|^2.
\end{eqnarray}
Here, the summation is over all the initial momenta $\vec{k}'$ leading to the final momentum $\vec{k}$.
$D$ is the matrix element emerging when the saddle-point method is applied to calculate the SFA ionization amplitude and $C_{\textrm{Coul}}$ is the Coulomb correction of the ionization rate \cite{PPT1967}. The phase associated with every trajectory is decomposed in Eq.~(\ref{scts_new}) as $S^{0}_{\downarrow}+S_{\rightarrow}$, where 
\begin{equation}
\label{s1}
S^{0}_{\downarrow}=I_{p}t_{s}-\frac{1}{2}\int_{t_s}^{t_0}dt\left[\vec{k}'+\vec{A}\left(t\right)\right]^{2}
\end{equation}
corresponds to the ionization step (motion under the potential barrier), and 
\begin{eqnarray}
\label{s2}
S_{\rightarrow}&=& \nonumber\\
&-&\int_{t_0}^\infty dt\, \left\lbrace\frac{p^2(t)}{2}+V[\vec{r}(t)]-\vec r(t)\cdot\vec\nabla V[\vec{r}(t)]\right\rbrace
\end{eqnarray}
accounts for the electron motion in the continuum. We note that coincides with the third term of Eq.~(\ref{phas_sim}).
The Jacobian $J$ is calculated as
\begin{equation}
\label{jac}
J\left(t\right)=\textrm{det}\left(\frac{\partial \vec{k}\left(t\right)}{\partial \vec{k}'}\right).
\end{equation}
The Maslov index $\nu$ changes at focal points, i.e, at times $T$ when the Jacobian is zero $J\left(T\right)=0$. The change (jump) of the Maslov index when the trajectory passes through a focal point is calculated as:
\begin{equation}
\label{jump}
\Delta\nu\left(T\right)=m-1+\textrm{sgn}~\textrm{det}\left(g\right), 
\end{equation}
where the $m\times m$ matrix $g$ is given by
\begin{equation}
\label{matrix_g}
g_{i,j}=\delta\vec{r}^{\left(i\right)}\cdot\textrm{Hesse}_{\vec{r},\vec{r}}\left(H\right)\delta\vec{r}^{\left(j\right)}
\end{equation} 
Here, in turn, $m$ is the number of linearly independent directions $\vec{d}^{\left(i\right)}$ $\left(i=1,...,m\right)$, which can be found at the focal points, such that infinitesimal changes of the initial momenta in these directions $\vec{k}'\rightarrow\vec{k}'+\epsilon\vec{d}^{\left(i\right)}$ do not affect $\vec{k}\left(T\right)$ in the first order of $\epsilon$. These changes of the initial momenta correspond to the changes of the position 
\begin{equation}
\label{chng}
\delta\vec{r}^{\left(i\right)}=\epsilon\sum_{j}\frac{\partial\vec{r}\left(T\right)}{\partial k'_{j}}d_{j}^{\left(i\right)},
\end{equation}
see Eq.~(\ref{matrix_g}). The Hessian $\textrm{Hesse}_{\vec{r},\vec{r}'}\left(H\right)$ of the Hamiltonian function 
\begin{equation}
\label{hamilt_vel}
H=\frac{1}{2}\left[\vec{k}+\vec{A}\left(t\right)\right]^2+V\left(\vec{r}\right)
\end{equation}
is calculated with respect to the position vector $\vec{r}$. 

The inverse problem is solved in Ref.~\cite{Brennecke2020} by using clustering algorithms. More specifically, density-based spatial clustering of applications with noise algorithms was applied. The solution of inverse problem with clustering shows an example of the application of machine learning (see Ref.~\cite{Raschka} for a text-book treatment) to strong-field phenomena. Other recent applications of the machine learning in strong-field physics are discussed in, e.g., Refs.~\cite{Yang2020,Lytova2021}. 

The fact that the Jacobian is explicitly taken into account in Eq.~(\ref{scts_new}) along with the solution of the inverse problem ensures the correct preexponential weight of every trajectory, namely, $1/\sqrt{\left|J\right|}$. It should be emphasized that this weight cannot be reproduced in ``shooting method", since the distribution of the trajectories over the cells in accord with their final momenta automatically creates a factor of $1/\left|J\right|$ instead of the $1/\sqrt{\left|J\right|}$. This problem was ignored in the implementation of the SCTS \cite{Shvetsov2016}, since the implementation of Ref.~\cite{Shvetsov2016} accounts only for the exponential factors in the trajectories weights.

The simple relation between the Jacobian in the 3D case and the corresponding Jacobian for two spatial dimensions was derived in \cite{Brennecke2020} for systems (ionic potential and the laser field) with cylindrical symmetry: 
\begin{equation}
\label{Jacobians2D3D}
\left|J_{3D}\right|=\frac{k_{\perp}}{k_{\perp}'}\left|J_{2D}\right|,
\end{equation}
where $k_{\perp}=\sqrt{k_x^{2}+k_{y}^2}$ (the field is polarized along the $z$-axis). This correction weight allows to obtain the results for the 3D system performing only the 2D simulations, and, by doing so, reduce the computational costs of the SCTS model significantly. We note that Eq.~(\ref{Jacobians2D3D}) has been already used in the SCTS simulations of Ref.~\cite{Shvetsov2016}. 

The modified version of the SCTS is in excellent agreement with solution of the TDSE. This applies for both electron momentum distributions and energy spectra \cite{Brennecke2020}. It is shown that the inclusion of the preexponential factors is crucial for quantitative agreement with the TDSE results. The extended version of the SCTS can be applied not only to the linearly polarized pulses, but also to non-cylindrically-symmetric laser \\ fields, e.g., bicircular ones, see Ref.~\cite{Brennecke2020}. Undoubtedly the version of the SCTS developed in \cite{Brennecke2020} is a valuable tool that is extremely useful in studies of strong-field ionization.

\section{Semiclassical two-step model and the strong-field holography with photoelectrons}

Development of the techniques capable to image the atomic positions that change in time in a chemical reaction will lead to a revolution in chemistry, biology, nanoscience, etc. At present there are many methods for time-resolved molecular imaging (see Ref.~\cite{Agostini2016} for a review). These methods have been developed due to the prominent progress in laser technologies. This applies above all to the development of the technology for pulse compression and the emergence of free-electron lasers. Moreover, the availability of table-top intense femtosecond lasers, which led to the emergence of strong-field, ultrafast, and attosecond physics, gave a strong impulse to the development of new techniques for time-resolved molecular imaging. Among these techniques are: laser-induced Coulomb-explosion imaging \cite{Frasinski1987,Cornaggia1991,Posthumus1995,Cornaggia1995}, laser-assisted electron diffraction \cite{Kanya2010,Morimoto2014}, high-order harmonic orbital tomography \cite{Itatani2004,Haessler2010}, laser-induced electron diffraction (see, e.g., Refs.~\cite{Meckel2008,Blaga2012,Pullen2015}), and strong-field photoelectron holography (SFPH)~\cite{Huismans2011}.

The SFPH method implements the widely-known idea of holography (1971 Nobel Prize in Physics awarded to Dennis Gabor, see Ref.~\cite{Gabor1971}) in strong-field physics. It was for the first time shown in 2011 by Y.~Huismans et al. \cite{Huismans2011} that a holographic pattern can be clearly recorded in experiment. This pattern in the electron momentum distributions is created by the signal (rescattered) and reference (direct) electrons. The SFPH can be implemented in a table-top experiment. It was shown that the holographic patterns encode a lot of spatio-temporal information about both the parent ion and the recolliding electron \cite{Huismans2011}. Last but not least, the electron dynamics can be imaged with subcycle (i.e., attosecond) time resolution. These advantages have triggered extensive studies of the SFPH, both experimental \cite{Marchenko2011,Hickstein2012,Meckel2014,Haertelt2016,Walt2017} and theoretical \cite{Huismans2011,Marchenko2011,Hickstein2012,Huismans2012,Bian2011,Bian2012,Bian2014,Li2014,Li2015,Faria2017a,Faria2017b,Faria2018aa,Faria2018a,Faria2018b}.

However, the first SFPH experiments \cite{Huismans2011,Marchenko2011,Hickstein2012,Huismans2012,Meckel2014} investigated the ionization process and the dynamics of the electron wave packet rather than molecular structure or dynamics. This is because of the fact that for diatomic and small molecules the holographic structures are mostly determined by the long-range and the alignment-independent Coulomb potential. As the result, the short-range effect reflecting the molecular structure cannot be observed on the background of the more intense Coulomb contribution. This problem was elegantly solved in experiment of Ref.~\cite{Haertelt2016} by considering the difference between the normalized photoelectron holograms for aligned and antialigned molecules.
This approach is based on the fact that for large scattering angles the differential cross section deviates from the Coulomb one and depends on the alignment of the molecule at the ionization instant. A similar method was also used in Ref.~\cite{Walt2017}. Various approaches were used for theoretical analysis of the SFPH: the three-step model \cite{Bian2011,Bian2012,Bian2014,Li2015}, the SFA version that accounts for rescattering \cite{Huismans2011,Huismans2012}, the Coulomb-corrected strong-field approximation \cite{Huismans2011,Huismans2012}, the CQSFA~\cite{Faria2017a,Faria2017b,Faria2018a,Faria2018b}, etc. (see Ref.~\cite{Faria2020} for recent review).  

\subsection{SCTS model and experimental holographic patterns}
 
\begin{figure}[h]
\begin{center}
\includegraphics[width=0.45\textwidth]{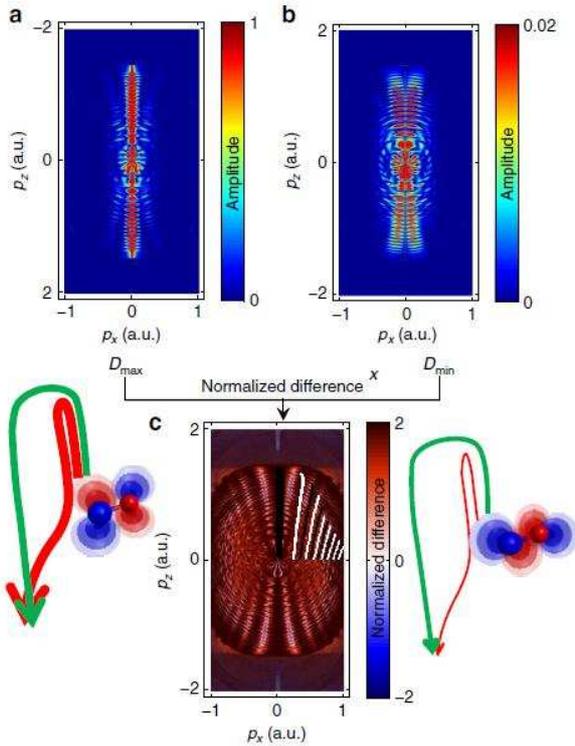} 
\end{center}
\caption{Photoelectron momentum distributions for ionization of the NO molecule by a laser pulse with a duration of 35~fs, intensity of $2.3\times10^{14}$ W/cm$^2$, and wavelength of 800 nm calculated using the SCTS model. The panels (a) and (b) show the distributions obtained in the cases where the electron density of the HOMO is aligned along the laser polarization direction and perpendicular to it, respectively. Panel (c) presents the normalized difference of the distributions shown in (a) and (b). The figure is reprinted from Ref.~\cite{Walt2017}.}
\label{fig6}
\end{figure}

The SCTS model was applied to the simulations of the holographic interference patterns observed in the experiment \cite{Walt2017}. In the study \cite{Walt2017} the electron momentum distribution produced in ionization of the NO molecule were calculated for two different cases. In the first case the electron density of the highest occupied molecular orbital (HOMO) is aligned along the polarization direction, whereas in the second case this density is orthogonal to it. These distributions, as well as their normalized difference are shown in Fig.~\ref{fig6}. To apply the SCTS model, the distributions over the initial transverse velocities are needed for both these cases. These distributions  were determined using the approach based on partial Fourier transform generalized to molecules (MO-PFT) \cite{Ivanov2010,Liu2016a,Liu2017}. The MO-PFT approach works with the electron wave function in mixed (coordinate-momentum) representation and uses the Wentzel-Kramers-Brillouin (WKB) approximation. The MO-PFT requires the corresponding HOMO's that were obtained using the GAMESS package \cite{Gamess}. The semiclassical simulations are in a perfect agreement with the experimental results \cite{Walt2017}. The simulations within the SCTS model reproduce all characteristic features of the holographic patterns. The regions of constructive and destructive interference predicted by the model of Ref.~\cite{Bian2011} that neglects the Coulomb potential are shown in Fig.~\ref{fig6} with white and black color, respectively. It is seen that the there-step model overestimates the spacing between the holographic fringes in the direction perpendicular to laser polarization. Therefore, the account of the Coulomb potential leads to the improved agreement between the experiment and the semiclassical simulations. 

\subsection{Effects of the Coulomb potential and the strong-field photoelectron holography}

The three-step semiclassical model predicts different types of subcycle interferometric structures, see Ref.~\cite{Bian2011}. Various types of the holographic structures arise due to the fact that the reference and signal electrons can start from different quarter cycles of the laser field.
In Ref.~\cite{Shvetsov2018a} the various types of the subcycle interference patterns revealed in \cite{Bian2011} were calculated accounting for the Coulomb potential of the ion with the adapted version of the SCTS model. Here we sketch the main points of this adapted SCTS. 

First, an ensemble of classical trajectories is launched only from the central period of a long ($8$ optical cycles) laser pulse. Second, the simple formula entirely neglecting the Coulomb potential, i.e., considering triangular potential barrier formed by the laser field and the ground state energy, is used for the tunnel exit point:
\begin{equation}
\label{trg}
\left|z_e\left(t_0\right)\right|=-\frac{I_p}{F\left(t_0\right)},
\end{equation}
where the sign of $z_e\left(t_0\right)$ is to be chosen to ensure the electron tunnels in the direction opposite to the instantaneous field $\vec{F}\left(t_0\right)$. This makes it possible to directly compare the resulting interference patterns with the patterns of the three-step model. Third, the weights (\ref{tunrate}) of classical trajectories were not taken into account, and the trajectories were distributed uniformly, which is justified by the fact that holographic patterns and not electron momentum distributions were calculated in Ref.~\cite{Shvetsov2018a}. Finally, a special approach instead of Eq.~(\ref{stad_int}) has to be used in the semiclassical model to obtain the phase difference between the signal and reference electrons. Indeed, to calculate the phase difference we need to isolate only one kind of rescattered trajectories and only one kind of the direct ones. This is a complicated task if the Newton’s equation of motion (\ref{newton}) is solved treating the laser field and the Coulomb force on equal footing. First of all, it is necessary to answer the question: How to distinguish between the direct and rescattered electron trajectories in the presence of the Coulomb field? Indeed, all the trajectories are, to some extent, affected by the Coulomb potential.

The following simple recipe is used in Ref.~\cite{Shvetsov2018a}. The reference trajectories were defined as those passing the ionic core at large distances and thus experiencing small-angle scattering only. More precisely, the reference electrons obey the condition $v_{0,\perp}k_{y}\geq0$. In contrast to them, the signal trajectories come close to the parent ion and undergo large-angle scattering that changes direction of the $k_y$ component compared to the initial one. Therefore, the signal trajectories can be defined as obeying the condition $v_{0,\perp}k_{y}\leq0$. However, these conditions are not sufficient to calculate the holographic structures correctly. The fact is that in the presence of the Coulomb field the mapping from the plane of initial conditions $\left(t_0,v_{0,\perp}\right)$ to the $\left(k_x,k_y\right)$ plane is a complicated function. For example, in the domain where the condition $v_{0,\perp}k_{y}\leq0$ defining the signal trajectories is fulfilled, this mapping is not one-to-one: Different sets of initial conditions lead to the same momentum $\vec{k}$, see Ref.~\cite{Shvetsov2018a} for details. The separation of trajectories of different kinds can be efficiently done by using the clusterization algorithms. In Ref.~\cite{Shvetsov2018a} this trajectory separation was accomplished manually by careful inspection of the mapping $\left(t_0,v_{0,\perp}\right)\rightarrow\left(k_x,k_y\right)$. 

\begin{figure}[h]
\begin{center}
\includegraphics[width=0.5\textwidth]{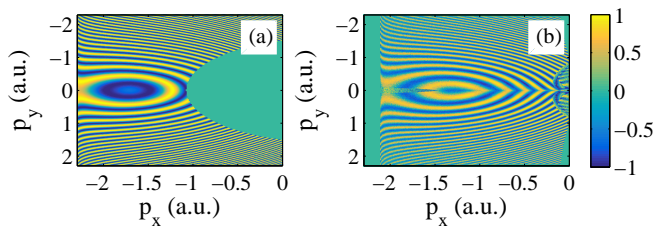} 
\end{center}
\caption{Holographic patterns emerging due to interference of a direct electron with a rescattered one that has the shortest travel time (see Ref.~\cite{Shvetsov2018a}) calculated (a) using the three-step model with time-dependent exit point, and (b) accounting for the Coulomb potential of the ion. The interference patterns are calculated for ionization of the H atom at a wavelength of 800 nm and intensity of $6.0\times10^{14}$ W/cm$^2$.}
\label{fig7}
\end{figure}

It was found that the Coulomb potential changes interference patterns significantly. Three main effects of the Coulomb field in the holographic patterns were identified in Ref.~\cite{Shvetsov2018a}. These are: shift of the interference pattern as a whole, filling of the parts of the pattern that are unfilled when the Coulomb potential is disregarded, and the characteristic kink of the interference pattern in the vicinity of $k_y=0$ [cf. Figs.~\ref{fig7}~(a) and (b)]. This kink at zero transverse momenta was attributed to the Coulomb focusing effect \cite{Ivanov1996}. However, the question remains, how sensitive are the predicted Coulomb effects to focal averaging. Therefore, further studies are required to understand which of these  effects can be observed in experiment. 

\section{Semiclassical two-step model and multielectron polarization effects}

The theoretical methods used in strong-field physics usually employ the single-active electron approximation (SAE). In the SAE an atom or molecule interacting with the laser pulse is replaced by a single electron. This single electron moves in the laser field and in the field of an effective potential. Therefore, the ionization is treated as a one-electron process. The SAE is a basis for understanding of many strong-field processes, including ATI and HHG \cite{BeckerRev2002,GrossmannBook}. Nevertheless, the role of the multielectron effects (ME) in strong-field and ultrafast physics has been attracting particular attention (see, e.g., Refs.~\cite{Kang2018,Le2018} and references therein). By now many theoretical approaches aimed at the description of the ME effects have been developed. The most well-known and widely used of them are: the time-dependent density-functional theory \cite{RungeGross1984} (see Refs.~\cite{UllrichBook,GrossBook} for a text-book treatment), multiconfiguration time-dependent Hartree-Fock theory \cite{Zanghellini2004,Caillat2005}, time-dependent restricted-active-space \cite{Madsen2013} and time-dependent complete-active-space self-consistent field theory \cite{Ishikawa2013}, time-dependent R-matrix theory \cite{Burke1997,Lysaght2009} and R-matrix theory with time-dependence \cite{Nikolopoulos2008,Moore2011}, time-dependent analytical R-matrix theory \cite{Torlina2012}, etc. (see Ref.~\cite{Shvetsov2018b}). There are also many semiclassical approaches capable to account for the ME effects, see, e.g. Refs.~\cite{Keller2012,Shvetsov2012,Dimitrovski2014,Dimitrovski2015,Dimitrovski2015JPB,Kang2018}. The advantages of the trajectory-based models discussed in Sec.~1 are particularly valuable in studies of complex ME effects.

One of the most well-known ME effects in strong-field ionization is laser-induced polarization of the parent ion. Recently the polarization effects in the ATI have been actively studied, see, e.g., Refs.~\cite{Dimitrovski2010,Keller2012,Shvetsov2012,Dimitrovski2014,Dimitrovski2015,Dimitrovski2015JPB,Kang2018}. In Refs.~\cite{Brabec2005,Zhao2007} and \cite{Dimitrovski2010} the effective potential for the outer electron that accounts for the external laser field, the Coulomb interaction, and the polarization effects of the ionic core, is derived in the adiabatic approximation. It was for the first time found in Ref.~\cite{Keller2012} that the time-independent Schrodinger equation with this effective potential and accounting for the Stark-shift of the ionization potential can be approximately separated in parabolic coordinates. This separation determines a certain tunneling geometry. The emerging physical picture of the flow of the electron charge associated with the tunneling electron is referred to as tunnel ionization in parabolic coordinates with induced dipole and Stark shift (TIPIS). The semiclassical model based on the TIPIS approach and disregarding the interference effect has shown a good agreement with experimental data (see Refs.~\cite{Keller2012,Dimitrovski2014,Dimitrovski2015JPB}) and the TDSE results \cite{Keller2012,Shvetsov2012}.

The electron momentum distributions generated in ionization of different atoms and molecules, including Ar, Mg, CO, naphthalene, etc., are very sensitive to the ME effects accounted by the induced dipole of the ionic core \cite{Keller2012,Shvetsov2012,Dimitrovski2014,Dimitrovski2015,Dimitrovski2015JPB}. These studies consider ionization by circularly or elliptically polarized laser pulses. This is due to the fact that the effective potential derived in Refs.~\cite{Brabec2005,Zhao2007} and \cite{Dimitrovski2010} is valid only at large and intermediate distances from the ionic core. In close to circularly polarized laser fields the rescattering-induced processes are suppressed (see Ref.~\cite{Dietrich1994}), and, therefore, the vast majority of the ionized electrons do not return to the parent ion. However, this is not true for linearly polarized field, and the applicability of the TIPIS approach in semiclassical simulations in the case of linear polarization raised questions. This problem is addressed in Ref.~\cite{Shvetsov2018b}. Furthermore, the study \cite{Shvetsov2018b} combines the TIPIS approach with the SCTS model. The resulting two-step semiclassical model for strong-field ionization is capable to describe quantum interference and accounts for the Stark-shift, the Coulomb potential, and the polarization induced dipole potential. 

\subsection{Combination of the TIPIS model and the STCS}
 
The ionic potential derived in Refs.~\cite{Brabec2005,Zhao2007,Dimitrovski2010} reads as:
\begin{equation}
\label{ME_pot}
V\left(\vec{r},t\right)=-\frac{Z}{r}-\frac{\alpha_{I}\vec{F}\left(t\right)\cdot\vec{r}}{r^3},
\end{equation}
where ME effect is accounted through the induced dipole potential $\left[\alpha_{I}\vec{F}\left(t\right)\cdot \vec{r} /r^3\right]$. For the potential of Eq.~(\ref{ME_pot}) the starting point of a classical trajectory can be obtained as the tunnel exit in the TIPIS model. More specifically, the tunnel exit point is given bt $z_e\approx-\eta_{e}/2$, where $\eta_{e}$ satisfies the equation:
\begin{equation}
\label{tunex2}
-\frac{\beta_{2}\left(F\right)}{2\eta}+\frac{m^2-1}{8\eta^2}-\frac{F\eta}{8}+\frac{\alpha_{I}F}{\eta^2}=-\frac{I_{p}\left(F\right)}{4},
\end{equation}
It is seen that Eq.~(\ref{tunex2}) has the additional ME term in the left-hand side compared to the equation (\ref{tunex}). Since the ME term in the potential (\ref{ME_pot}) is proportional to the laser field $\vec{F}(t)$, it is absent at $t>t_f$. Therefore, after the laser pulse terminates, the electron moves in the Coulomb field only. This makes it possible to use Eq.~(\ref{mominf}) for calculation of the asymptotic momentum of the electron from its position and momentum at $t=t_f$. The SCTS phase (\ref{phas_sim}) with the potential $V\left(\vec{r},t\right)$ defined by Eq.~(\ref{ME_pot}) reads as:
\begin{eqnarray}
\label{phas_me}
& & \Phi^{SCTS}\left(t_{0},\vec v_0\right)= - \vec v_0\cdot\vec r(t_0) + I_{p}t_{0} \nonumber\\
&-& \int_{t_0}^\infty dt\, \left\lbrace\frac{p^2(t)}{2}-\frac{2Z}{r}-\frac{3\alpha_{I}\vec{F}\left(t\right)\cdot \vec{r}}{r^3}\right\rbrace\ .
\end{eqnarray}
In order to implement the resulting semiclassical model, the importance sampling method was used in Ref.~\cite{Shvetsov2018b}. We note that in addition to the inapplicability of the potential (\ref{ME_pot}) at small distances, there exist other conditions that restrict the range of applicability of the TIPIS model (see Refs.~\cite{Shvetsov2012,Shvetsov2018b} for details). The study \cite{Shvetsov2018b} focuses on the cases of Mg ($I_p=0.28$~a.u., $\alpha_N=71.33$~a.u., $\alpha_{I}=35.00$~a.u.) and Ca ($I_p=0.22$~a.u., $\alpha_N=169.0$~a.u., $\alpha_{I}=74.11$~a.u.) atoms, which have similar ionization potentials. But their static ionic polarizabilities are different by approximately two times.  

In order to avoid the application of the potential (\ref{ME_pot}) at small distances, a special cutoff radius $r_C$ was introduced in Ref.~\cite{Shvetsov2018b}, and all the trajectories entering the sphere $r<r_{C}$ were ignored. The remaining trajectories do not reach the vicinity of the ion. It is clear that the elimination of the whole class of the trajectories (the returning ones) depletes some parts of electron momentum distributions. However, these depleted parts usually correspond to the boundary of the direct ionization spectrum. Therefore, they do not affect the main part of the momentum distributions that provides major contribution to the ionization yield, see Ref.~\cite{Shvetsov2018b} for details.  

\subsection {Application of the combined semiclassical model}

\begin{figure}[h]
\begin{center}
\includegraphics[width=0.5\textwidth]{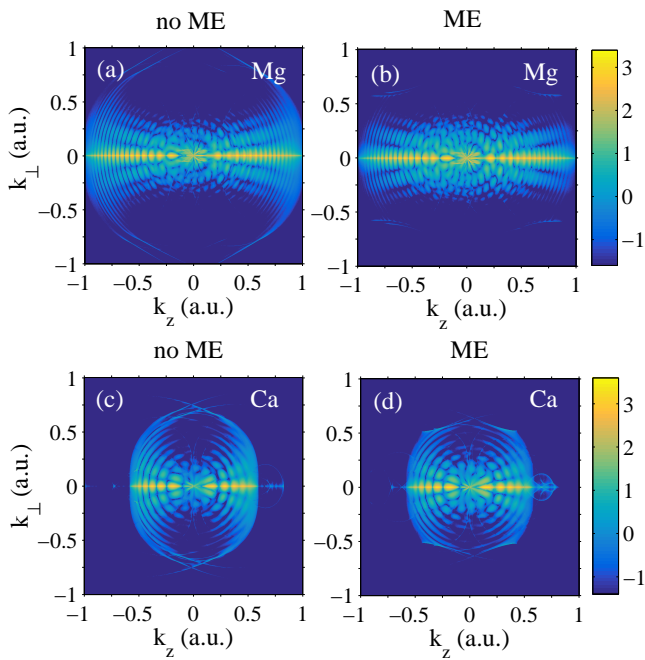} 
\end{center}
\caption{Two-dimensional electron momentum distributions for the Mg [(a),(b)] and Ca [(c),(d)] atoms calculated by combining the TIPIS approach with the SCTS model. The wavelength is 1600 nm and the pulse duration is $n=8$ cycles. Panels [(a),(b)] and [(c),(d)] show the distributions calculated at the intensities of $3.0\times10^{13}$~W/cm$^2$ and $1.0\times10^{13}$~W/cm$^2$, respectively. The distributions [(a), (c)] are obtained neglecting the ME terms in Eqs.~(\ref{ME_pot}), (\ref{tunex2}), and (\ref{phas_me}), whereas the distributions [(b),(d)] are calculated accounting the ME terms in all these equations. The momentum disributions are normalized to the total ionization yield. A logarithmic color scale in arbitrary units is used.}
\label{fig8}
\end{figure}

The 2D photoelectron momentum distributions calculated in accord with the resulting semiclassical model are shown in Figs.~\ref{fig8}~(a)-(d). Figures \ref{fig8}~(a) and (c) correspond to the distributions calculated accounting for the laser and Coulomb fields. Figures \ref{fig8}~(b) and (d) display the results of the combined TIPIS + SCTS model, i.e., with the account of the ME potential. The panels [(a), (b)] and [(c), (d)] correspond to ionization of Mg and Ca, respectively. It is seen that the presence of the ME term in the potential of Eq.~(\ref{ME_pot}) results in a narrowing of the longitudinal momentum distributions and modification of the interference structures.

\begin{figure}[h]
\begin{center}
\includegraphics[width=0.5\textwidth]{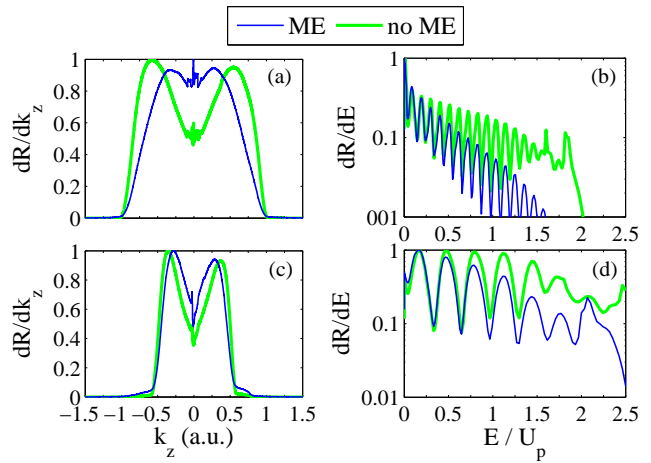} 
\end{center}
\caption{(a),(c) Electron momentum distributions in the longitudinal direction, and (b),(d) energy spectra. Panels (a,b) and (c,d) correspond to the ionization of Mg and Ca, respectively. Thick green curve and thin blue curve show the semiclassical results obtained with and without ME terms, respectively. The wavelength and duration of the pulse are as in Fig.~\ref{fig8}. The panels (a,b) and (c,d) are calculated for the intensities of $3.0\times10^{13}$~W/cm$^2$ and $1.0\times10^{13}$~W/cm$^2$, respectively. The longitudinal distributions and energy spectra are normalized to the maximum value.}
\label{fig9}
\end{figure}

We first discuss the narrowing of the longitudinal distributions. This effect is further illustrated in Figs.~\ref{fig9} (a) and (c) that show the longitudinal momentum distributions obtained with and without the ME term for Mg and Ca, respectively. Since the widths of the distributions do not change due to the interference effects, the phase is disregarded in the calculations of Figs.~\ref{fig9}~(a) and (c). The corresponding electron energy spectra are shown in Figs.~\ref{fig9}~(b) and (d). It is seen that the spectra calculated accounting for the ME term fall off more rapidly with increase of the energy than the ones obtained neglecting the ME effects. This is a direct consequence of the narrowing of the corresponding 2D electron momentum distributions. 

The mechanism underlying the narrowing effect has a kinematic origin \cite{Shvetsov2018b}. The analysis of classical trajectories has shown that there is a certain class of trajectories strongly affected by the induced polarization of the ionic core. The trajectories of this class start closer to the parent ion that other trajectories and their initial transverse velocities are not too large (see Ref.~\cite{Shvetsov2018b} for details). Indeed, the force acting on the electron due to the ME polarization effect (the ME force) decays as $1/r^2$ with increasing $r$. Therefore, this force can change the electron motion only at the initial part of the trajectory adjacent to the tunnel exit. The ME force reduces both longitudinal and transverse components of the electron final momentum, and, as the result, the trajectories belonging to this class lead to the bins with smaller $\vec{k}$. We note that for close to circularly polarized laser pulses the ME effects result in the rotation of the 2D electron momentum distributions towards the small axis of polarization ellipse \cite{Keller2012}.

It is seen that the presence of the ME term in the equations of motion and the phase does not dramatically change the interference patterns. The interference structure is modified only in the first and the second ATI peaks and also in the vicinity of the $k_z$ axis. The analysis of the mechanism behind the polarization-induced interference effect showed that the changes in interference patterns are mostly caused by the ME term in the equation of motion, whereas the presence of the term $-3\alpha_{I}\vec{F}\cdot\vec{r}/r^3$ in the phase (\ref{phas_me}) does not play a substantial role \cite{Shvetsov2018b}.

\begin{figure}[h]
\begin{center}
\includegraphics[width=0.48\textwidth, trim={0 1cm 0 0.5cm}]{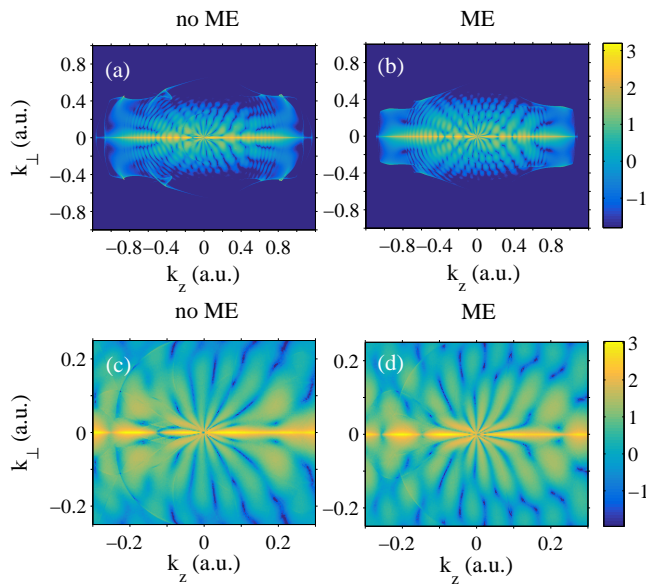} 
\end{center}
\caption{Two-dimensional photoelectron momentum distributions for the Ba atom ionized by a laser pulse with a duration of $n=4$ cycles, intensity of $3.0\times10^{13}$ W/cm$^2$ and a wavelength of 1600 nm obtained by semiclassical simulations neglecting the ME term in the phase (\ref{phas_me}) [(a),(c)] and including this term [(b),(d)]. Panels (c) and (d) display the magnification for $\left|k_z\right|\le0.3$~a.u. and $\left|k_{\perp}\right|\le0.25$~a.u. of the momentum distributions shown in (a) and (b), respectively. In both cases the ME force is included in the Newton's equation of motion. The normalization to the total ionization yield is used. The color scale is logarithmic with arbitrary units.} 
\label{fig10}
\end{figure}

It was found that the trajectories interfering in a given bin often have similar ME contributions to the phase 
\begin{equation}
\label{me_int}
-\int_{t_0}^{\infty}dt\frac{3\alpha_{I}\vec{F}\cdot\vec{r}\left(t\right)}{r^3\left(t\right)},
\end{equation}
and therefore, the difference of these contributions is small. This difference is the only important quantity for the interference effect. The ME contributions to the phase are similar due to the combination of the following reasons: (i) the tunneling probability is a sharp function of the laser field $F\left(t_0\right)$ at time of ionization, (ii) the tunnel exit depends only on $\vec{F}\left(t_0\right)$ and the parameters of the atom (molecule), and (iii) only the initial part of the electron trajectory is relevant in the integral (\ref{me_int}). Nevertheless, for atoms and molecules with large values of the ionic polarizability $\alpha_I$, the difference of the ME contributions to the phase is essential. As the result, the changes in the interference patters due to the ME effect can be significant. This is illustrated in Figs.~\ref{fig10}~(a)-(d). It is seen that the number of radial nodal lines in the fanlike interference pattern at low energies is different when calculated with and without the ME term in the phase of Eq.~(\ref{phas_me}). In Fig.~\ref{fig10}~(c) there are six nodal lines for positive $k_{\perp}$, while in the presence of the ME term only five such lines are visible [see Fig.~\ref{fig10}~(d)]. 

\section{Semiclassical two-step model for H$_2$ molecule}

To the best of our knowledge, there are only a few works that apply semiclassical models accounting for the quantum interference effect to describe strong-field ionization of molecules, see Refs.~\cite{Liu2016,Liu2017,Shvetsov2019a}. The studies \cite{Liu2016,Liu2017} extend the QTMC model to the molecular case. The SCTS model was applied to the hydrogen molecule in Ref.~\cite{Shvetsov2019a}. Two-dimensional electron momentum distributions, energy spectra, and angular distributions were compared to the ones calculated for ionization of the atomic hydrogen. The study \cite{Shvetsov2019a} revealed substantial differences in electron momentum distributions and energy spectra as compared to the atomic case. 

\subsection{SCTS model for hydrogen molecule}
The ionic potential experienced by a single-active-electron in the H$_2$ molecule is given by
\begin{equation}
\label{pot_H2}
V\left(\vec{r}\right)=-\frac{Z_1}{\left|\vec{r}-\vec{R}/2\right|}-\frac{Z_2}{\left|\vec{r}+\vec{R}/2\right|}
\end{equation}
Here $\vec{R}$ is the vector pointing from one nucleus to another. It is assumed that the origin of the coordinate system is located in the center of the molecule. The effective charges $Z_1$ and $Z_2$ are chosen to be equal to $0.5$~a.u. \cite{Liu2016,Liu2017}. It is obvious that the question how to distribute initial conditions of classical trajectories is more complicated for a molecule than for an atom. Presently there are two well-known approaches to this problem: Molecular quantum-trajectory Monte-Carlo model (MO-QTMC) \cite{Liu2016,Liu2017} and MO-PFT \cite{Ivanov2010,Liu2016a,Liu2017}. The MO-QTMC model applies expressions of the molecular strong-field approximation (MO-SFA) \cite{Faisal2000,Madsen2004}. The MO-SFA is a generalization of the SFA that was initially developed for atoms to the case of molecules. The MO-PFT model was used in Ref.~\cite{Shvetsov2019a}.   

The bound state orbital in the H$_2$ molecule is the bonding superposition of the two 1s atomic orbitals located at the centers of the atoms:
\begin{align}
\label{Homo_H2}
\Psi_{H_{2}}\left(\vec{r}\right)&=\frac{1}{\sqrt{2\left(1+S_{OI}\right)}}\left[\psi_{atom1}\left(\vec{r}-\vec{R}/2\right)\right.\nonumber \\
&+\left.\psi_{atom2}\left(\vec{r}+\vec{R}/2\right)\right].
\end{align} 
The corresponding partial Fourier transform is given by (see Ref.~\cite{Liu2016a}):\\ 
\begin{eqnarray}
\label{part_H2}
& & \Pi_{H_2}\left(p_x,p_y,z\right)=\exp \left(-\frac{i}{2}R\sin\theta_{m}\left[p_x\cos\varphi_{m}\right.\right. \nonumber \\
&+& \left.\left.p_y\sin\varphi_{m}\right]\vphantom{\frac{1}{2}}\right)\Pi_{atom1}\left(p_x,p_y,z-\frac{R}{2}\cos\theta_{m}\right) \nonumber\\
&+& \exp \left(\frac{i}{2}R\sin\theta_{m}\left[p_x\cos\varphi_{m}\right.\right. \nonumber \\
&+& \left.\left.p_y\sin\varphi_{m}\right]\vphantom{\frac{1}{2}}\right)\Pi_{atom2}\left(p_x,p_y,z+\frac{R}{2}\cos\theta_{m}\right).
\end{eqnarray}
Here $\theta_{m}$ and $\varphi_{m}$ are the polar and azimuthal angles of the molecular axis, respectively, and $\Pi_{atom}\left(p_x,p_y,z\right)$ is the partial Fourier transform of the $1s$ orbital. Substituting the expression for $\Pi_{atom}\left(p_x,p_y,z\right)$ (see Ref.~\cite{Ivanov2010}) in Eq.~(\ref{part_H2}) we obtain the following formula for the mixed-representation wave function of the H$_2$ molecule applicable just beyond the tunnel exit:
\begin{eqnarray}
\label{distr}
& &\Pi\left(p_{x},p_{y},z_{e}\right)\nonumber \\
& & \sim\left\{\exp\left(-\frac{i}{2}R\sin\theta_{m}\left[p_x\cos\varphi_{m}+p_y\sin\varphi_m\right]\right)\right.\nonumber\\
& & \left.\times\exp\left(-\frac{1}{2}\kappa R \cos \theta_{m} \right)\nonumber\right. \nonumber\\
&+& \left.\exp\left(\frac{i}{2}R\sin\theta_{m}\left[p_x\cos\varphi_{m}+p_y\sin\varphi_m\right]\right)\right. \nonumber\\
& & \left.\times\exp\left(-\frac{1}{2}\kappa R \cos \theta_{m} \right)\right\} \nonumber\\
& & \times \exp\left[-\frac{\kappa^3}{3F}-\frac{\kappa\left(p_{x}^2+p_{y}^2\right)}{2F}\right].
\end{eqnarray}
As in Ref.~\cite{Liu2017}, this expression (without prefactor) was used in \cite{Shvetsov2019a} as a complex amplitude describing ionization at time $t_0$ with initial transverse velocity $v_{0,\perp}=p_{0,\perp}$. In the simplest case analyzed in Ref.~\cite{Shvetsov2019a} the molecule is oriented along the laser polarization direction $\left(\theta_m = \varphi_m = 0 \right)$, and the factor in brackets in Eq.~(\ref{distr}) is constant for a fixed internuclear distance $R$. This allows to use only the exponential factor of Eq.~(\ref{distr}). 

Different approaches can be used to find the tunnel exit point, i.e., the starting point of the trajectory, in the molecular case. The simplest one consists in neglecting the molecular potential, i.e., considering triangular potential barrier (\ref{trg}). An alternative approach, the so-called field direction model (DFM) (see Ref.~\cite{Shvetsov2012}), accounts for the molecular potential. The potential barrier in the FDM model is formed by the molecular potential and the laser field in a 1D cut along the field direction. Therefore, the tunnel exit point in the FDM model is defined by the equation:   
\begin{equation}
\label{fdm}
V\left(\vec{r}\right)+F\left(t_0\right)z_{e}=-I_{p}.
\end{equation}
To finalize the generalization of the SCTS model to the case of the H$_2$ molecule, we need to obtain the phase, which is assigned to a classical trajectory. This phase is derived by substituting the potential (\ref{pot_H2}) in Eq.~(\ref{phas_sim}):
\begin{align}
\label{phas_mol}
& \Phi_{H_{2}}^{SCTS}\left(t_{0},\vec v_0\right)= - \vec v_0\cdot\vec r(t_0) + I_{p}t_{0} \nonumber\\
& -\int_{t_0}^\infty dt\ \left\lbrace\frac{p^2(t)}{2}-\frac{Z_{1}\left(\vec{r}-\vec{R}/2\right)\cdot\left(2\vec{r}-\vec{R}/2\right)}{\left|\vec{r}-\vec{R}/2\right|^3}\right. \nonumber\\
&\left.+\frac{Z_{2}\left(\vec{r}+\vec{R}/2\right)\cdot\left(2\vec{r}+\vec{R}/2\right)}{\left|\vec{r}+\vec{R}/2\right|^3}\right\rbrace\ ,
\end{align}
see Ref.~\cite{Shvetsov2019a}. 
It is seen that for $r\gg R/2$ this phase corresponds to the SCTS phase for the Coulomb potential $-Z/r$ with the effective charge $Z=Z_1+Z_2$.  In contrast to this, the QTMC phase for the H$_2$ molecule is given by:
\begin{align}
\label{QTMC_H2}
&\Phi_{H_{2}}^{QTMC}\left(t_{0},\vec v_0\right)= - \vec v_0\cdot\vec r(t_0) + I_{p}t_{0} \nonumber\\
& -\int_{t_0}^\infty dt \left\{\frac{p^2\left(t\right)}{2}-\frac{Z_{1}}{\left|\vec{r}-\vec{R}/2\right|}-\frac{Z_{2}}{\left|\vec{r}+\vec{R}/2\right|}\right\}.
\end{align} 
The expression \ref{phas_mol} can be simplified at large distances and, as the result, the SCTS phase for the H atom is reproduced. Finally, it is assumed in Ref.~\cite{Shvetsov2019a} that at the end of the laser pulse the ionized electron is far enough from both nuclei, i.e., $r\left(t_f\right) \gg R$. If this condition is met, after the end of the pulse the electron moves in the Coulomb field with the effective charge $Z$. Therefore, its asymptotic momentum can be calculated from Eq.~(\ref{mominf}), and the post-pulse phase is determined by Eq.~(\ref{post_pulse}). 

\subsection{Application of the SCTS model to H$_2$ molecule}

\begin{figure}[h]
\begin{center}
\includegraphics[width=0.4\textwidth]{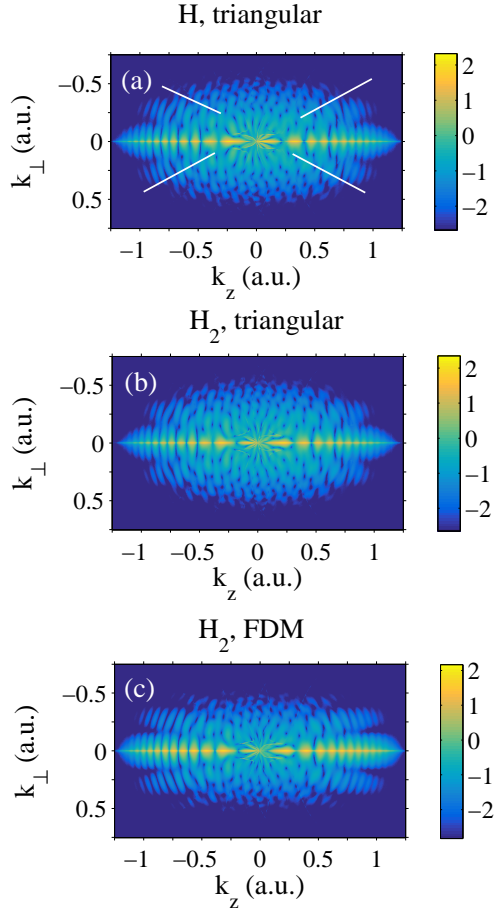} 
\end{center}
\caption{Two-dimensional electron momentum distributions for ionization of (a) the H atom and (b),(c) the H$_2$ molecule by a laser pulse with the duration of $n=4$ cycles, intensity of $2.0\times10^{14}$ W/cm$^2$, and wavelength of 800 nm. The distributions shown in panels (a) and (b) correspond to the tunnel exit point calculated from Eq.~(\ref{trg}). The distribution of panel (c) is obtained using the FDM expression for the tunnel exit.  The H$_2$ molecule is oriented along the polarization direction of the laser field ($z$-axis). The holographic fringes are shown by white lines in panel (a). The normalization to the total ionization yield is used. The color scale is logarithmic with arbitrary units.}  
\label{fig11}
\end{figure}

\begin{figure}[h]
\begin{center}
\includegraphics[width=0.45\textwidth]{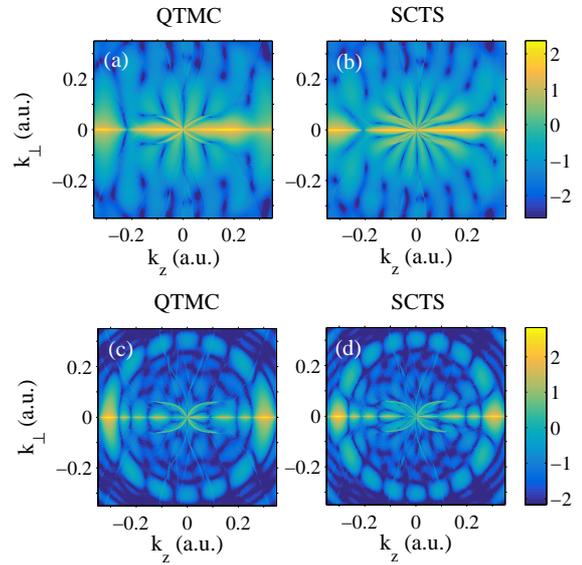} 
\end{center}
\caption{The low-energy parts of the two-dimensional photoelectron momentum distributions for the H$_2$ molecule ionized by a laser pulse with a duration of $n=4$ cycles, wavelength of 800 nm, and peak intensity of $1.2\times10^{14}$ W/cm$^2$. The left column [panels (a) and (c)] show the results of the QTMC model. The right column [panels (b) and (d)] present the distributions calculated within the SCTS model. Panels (a,b) and (c,d) are calculated for the sine squared and trapezoidal envelopes of the laser pulse, respectively (see Ref.~\cite{Shvetsov2019a}). The molecule is oriented along the laser polarization direction ($z$-axis). A logarithmic color scale in arbitrary units is used.}
\label{fig12}
\end{figure}

In Fig.~\ref{fig11} we compare the photoelectron momentum distributions calculated within the SCTS model for the hydrogen atom [Fig. \ref{fig11}(a)] and hydrogen molecule [Figs. \ref{fig11} (b) and \ref{fig11}~(c)], see Ref.~\cite{Shvetsov2019a}. The starting point of the trajectory for H is calculated using the triangular potential barrier (\ref{trg}). The distribution of Fig.~\ref{fig11}~(a) for H$_2$ is also obtained for the exit point calculated from Eq~(\ref{trg}). We note that the molecular potential is fully taken into account in the classical equations of motion (\ref{newton}) and in the phase (\ref{phas_sim}) when calculating Fig.~\ref{fig11}~(a). The electron momentum distribution of Fig.~\ref{fig11}~(a) corresponds to the tunnel exit obtained by using the FDM model. The electron momentum distributions shown in Figs.~\ref{fig11}~(a) and \ref{fig11}~(b) are similar to each other. Therefore, it can be concluded that if the molecular potential is not accounted in calculating the starting point, the effects of the molecular structure are not visible in electron momentum distributions. This result can be expected bearing in mind that $r_{0}=I_{p}/F_{0} \gg R/2$ for the parameters of Fig.~\ref{fig11}, and the distance between the ionized electron and the molecular ion increases further when the electron moves along the trajectory. As the result, the departing electron feels only the Coulomb asymptotic instead of the full molecular potential. The FDM model predicts smaller exit points as compared to the triangular barrier formula (\ref{trg}), see Ref.~\cite{Shvetsov2019a}. For this reason, the effects of the molecular potential (\ref{pot_H2}) are visible in Fig.~\ref{fig11}~(c). First, the photoelectron momentum distribution is more extended in the polarization direction. As the result, the energy spectra for the hydrogen molecule falls off slower with the increase of energy than the ones for the H atom. Simultaneously, the angular distributions in the molecular case are more aligned along the polarization direction. Second, at the same parameters of the laser pulse the holographic interference fringes are more pronounced for H$_2$ than for H (see Fig.~\ref{fig11}). 
The comparison of the distributions calculated using the SCTS and QTMC models for ionization of the H$_2$ molecule is presented in Figs.~\ref{fig12} (a)-(d). Two different pulse envelopes were used in Figs.~\ref{fig12}~(a)-(d). Figures \ref{fig12} (a) and (b) correspond to the sine squared pulse, whereas Figures \ref{fig12}~(c) and (d) show the distributions obtained for the trapezoidal pulse (see Ref.~\cite{Shvetsov2019a} for details). Figures~\ref{fig12}(a) and (c) display the momentum distributions calculated within the QTMC model, and Figs.~\ref{fig12}~(b) and (d) show the corresponding SCTS results. For the sine squared pulse these distributions have a pronounced fan-like structure in their low-energy part. For the trapezoidal envelope the fans are substituted by the characteristic blobs [see Figs.~\ref{fig12}(c) and (d)] lying on a circle with the radius $k=0.30$~a.u. Similar to the atomic case, the QTMC predicts fewer nodal lines in the interference structure at low energies than the SCTS model. This fact can be again attributed to the underestimation of the Coulomb potential in the QTMC phase \cite{Shvetsov2019a}.

\section{Conclusions}
The semiclassical models using classical mechanics to describe the electron motion after it has been released from an atom or molecule are one of the powerful methods of strong-field, ultrafast, and attosecond physics. The standard formulation of the trajectory-based models does not allow to describe the effects of quantum interference. Nevertheless, a substantial progress in simulations of the interference effects using the semiclassical models has been achieved recently. By present several trajectory-based models capable to describe the interference effects have been developed and successfully applied to the studies of the ATI. Here we discuss one of these models, namely, the SCTS. 

The SCTS model allows to reproduce interference patterns of the ATI process and accounts for the ionic potential beyond the semiclassical perturbation theory. In the SCTS the phase assigned to every classical trajectory is calculated using the semiclassical expression for the matrix element of the quantum mechanical propagator \cite{Miller1974,Walser2003,Spanner2003}. As the result, the SCTS model yields a good agreement with the direct numerical solution of the TDSE, better than, e.g., the QTMC model applying the first order semiclassical perturbation theory to account for the Coulomb potential in the phase.

Here we review further developments and applications of the SCTS. At first, we review the formulation of the SCTS and its numerical implementation. The application of the model was illustrated in the case of the H atom. We next turn to the further developments of the SCTS: the SCTSQI model \cite{Shvetsov2019b} and the SCTS model with the prefactor \cite{Brennecke2020}. In the SCTSQI model the initial conditions for classical trajectories are determined from the exact quantum dynamics of the wavepacket. For ionization of the 1D atom the SCTSQI model yields not only qualitative, but also quantitative agreement with the numerical solution of the TDSE. Further work is needed to accomplish the generalization of the SCTSQI on the 3D case. The developments in this direction have already begun. The quantitative agreement with the TDSE was also achieved by the extension of the SCTS model that accounts for the prefactor of the semiclassical matrix element. Furthermore, the 3D implementation of the SCTS \cite{Brennecke2020} has a number of other important modifications.

We discuss the application of the SCTS approach to the SFPH. The semiclassical simulations within the SCTS model are in perfect agreement with the results of the recent experiment \cite{Walt2017}. The model is able to reproduce all characteristic features of the observed holographic patterns. The SCTS model also allows to investigate the effect of the Coulomb potential on the holographic structures. Three main Coulomb effects in the interference patterns were predicted \cite{Shvetsov2018a}. However, it should be investigated how sensitive are these Coulomb effects to focal averaging. This further work will allow to understand, which of the predicted effects can be observed.

We also present a quick review of the application of the SCTS to study of the multielectron polarization effects. We discuss the modification of the SCTS model accounting for the multielectron polarization-induced dipole potential. The semiclassical simulations predict narrowing of the electron momentum distributions along the polarization direction. This narrowing arises due to the focusing of the ionized electrons by the induced dipole potential. Furthermore, the polarization of the ionic core can also modify the interference patterns in electron momentum distributions. 

Finally, we briefly reviewed the extension of the SCTS model to ionization of the hydrogen molecule. The SCTS model for the H$_2$ can be generalized to an arbitrary laser polarization and orientation of the molecule, as well as to heteronuclear and polyatomic molecules. We believe that these generalizations being combined with the extended versions of the SCTS will result to an emergence of powerfool tools for studies of the strong-field processes. 

\section*{Acknowledgements}
\noindent We are grateful to M.~Lein, L.~B.~Madsen, J.~Burgd\"orfer, H.~J.~W\"orner, C.~Lemell, D.~G.~Arb\'o, E.~R\"as\"anen, and K.~T\ifmmode \mbox{\H{o}}\else \H{o}\fi{}k\'esi for fruitful collaboration that resulted in some of the works discussed in this colloquium paper. We would also like to thank S.~Brennecke, N.~Eicke, C.~Faria, A.~Landsman, H.~Ni, F.~Oppermann, J.~Solanp\"a\"a, S.~Yue, and B.~Zhang for valuable discussions. This work was supported by the Deutsche Forschungsgemeinschaft (Grant No. SH 1145/1-2).

\end{document}